\documentclass[aps,prd,preprint,superscriptaddress,tightenlines,nofootinbib]{revtex4}
\usepackage{epsfig}
\usepackage{graphicx}
\usepackage{dcolumn}
\usepackage{bm}

\def\PI{\relax\ifmmode{\pi}\else{$\pi$}\fi}
\def\PIP{\relax\ifmmode{\pi^+}\else{$\pi^+$}\fi}
\def\PIZ{\relax\ifmmode{\pi^0}\else{$\pi^0$}\fi}
\def\PIM{\relax\ifmmode{\pi^-}\else{$\pi^-$}\fi}
\def\PIPM{\relax\ifmmode{\pi^\pm}\else{$\pi^\pm$}\fi}
\def\K{\relax\ifmmode{K}\else{$K$}\fi}
\def\KP{\relax\ifmmode{K^+}\else{$K^+$}\fi}
\def\KM{\relax\ifmmode{K^-}\else{$K^-$}\fi}
\def\KPM{\relax\ifmmode{K^\pm}\else{$K^\pm$}\fi}
\def\KMP{\relax\ifmmode{K^\mp}\else{$K^\mp$}\fi}
\def\D{\relax\ifmmode{D}\else{$D$}\fi}
\def\DB{\relax\ifmmode{\overline{D}}\else{$\overline{D}$}\fi}
\def\DZ{\relax\ifmmode{D^0}\else{$D^0$}\fi}
\def\DZB{\relax\ifmmode{\overline{D}^0}\else{$\overline{D}^0$}\fi}
\def\DP{\relax\ifmmode{D^+}\else{$D^+$}\fi}
\def\DM{\relax\ifmmode{D^-}\else{$D^-$}\fi}
\def\Ds{\relax\ifmmode{D_s^+}\else{$D_s^+$}\fi}
\def\RH{\relax\ifmmode{\rho}\else{$\rho$}\fi}
\def\RHP{\relax\ifmmode{\rho^+}\else{$\rho^+$}\fi}
\def\RHZ{\relax\ifmmode{\rho^0}\else{$\rho^0$}\fi}
\def\DS{\relax\ifmmode{D^\star}\else{$D^\star$}\fi}
\def\DST{\relax\ifmmode{D^*}\else{$D^*$}\fi}
\def\DSB{\relax\ifmmode{\overline{D}^\star}
    \else{$\overline{D}^\star$}\fi}
\def\DSZ{\relax\ifmmode{D^{\star0}}\else{$D^{\star0}$}\fi}
\def\DSZB{\relax\ifmmode{\overline{D}^{\star0}}
    \else{$\overline{D}^{\star0}$}\fi}
\def\DSP{\relax\ifmmode{D^{\star+}}\else{$D^{\star+}$}\fi}
\def\DSM{\relax\ifmmode{D^{\star-}}\else{$D^{\star-}$}\fi}
\def\DSPM{\relax\ifmmode{D^{\star\pm}}\else{$D^{\star\pm}$}\fi}
\def\DSMP{\relax\ifmmode{D^{\star\mp}}\else{$D^{\star\mp}$}\fi}
\def\B{\relax\ifmmode{B}\else{$B$}\fi}
\def\BB{\relax\ifmmode{\overline{B}}\else{$\overline{B}$}\fi}
\def\USSSS{\relax\ifmmode{\Upsilon({\rm 4S})}
    \else{$\Upsilon({\rm 4S})$}\fi}
\def\ee{\relax\ifmmode{e^+e^-}\else{$e^+e^-$}\fi}
\def\qq{\relax\ifmmode{q\overline q}\else{$q\overline q$}\fi}
\def\eecc{\relax\ifmmode{e^+e^-\rightarrow c\overline c}
    \else{$e^+e^-\rightarrow c\overline c$}\fi}
\def\PRD{Phys. Rev.}
\newcommand{\CL}{confidence level}

\def\chisq{\relax\ifmmode{\chi^2}\else{$\chi^2$}\fi}
\def\LUM{\relax\ifmmode{\cal{L}}\else{$\cal{L}$}\fi}

\def\fb{fb$^{-1}$}
\def\BR{\relax\ifmmode{\cal B}\else{$\cal B$}\fi}
\def\SIG{\relax\ifmmode{\sigma}\else{$\sigma$}\fi}
\def\TO{\relax\ifmmode{\rightarrow}\else{$\rightarrow$}\fi}
\def\Sim{\relax\ifmmode{\sim}\else{$\sim$}\fi}
\def\PM{\relax\ifmmode{\pm}\else{$\pm$}\fi}
\def\MC{Monte Carlo}
\def\MCS{Monte Carlo simulation}
\def\xp{\relax\ifmmode{x_p}\else{$x_p$}\fi}
\def\etal{{\textit et al.}}
\def\ie{\textit{i.e.}}
\def\eg{\textit{e.g.}}
\def\mdc{\relax\ifmmode{M_{cand}}\else{$M_{cand}$}\fi}
\def\mdsc{\relax\ifmmode{M^*_{cand}}\else{$M^*_{cand}$}\fi}
\def\EFFDZ{\relax\ifmmode{\epsilon_{D^0}}\else{$\epsilon_{D^0}$}\fi}
\def\EFFDP{\relax\ifmmode{\epsilon_{D^+}}\else{$\epsilon_{D^+}$}\fi}
\def\EFFDSP{\relax\ifmmode{\epsilon_{D^{*+}}}\else{$\epsilon_{D^{*+}}$}\fi}
\def\WRAT{\relax\ifmmode{\sigma_2/\sigma_1}\else{$\sigma_2/\sigma_1$}\fi}
\def\Journal#1,#2,#3,#4{{#1} {\bf #2}, #3 #4}

\begin{document}
\begin{flushright}
\preprint{CLNS 04/1861 v2}       
\preprint{CLEO 04-3 v2}         
\end{flushright}

\title{Charm meson spectra in \ee\ annihilation at 10.5~GeV c.m.e.}
\author{M.~Artuso}
\author{C.~Boulahouache}
\author{S.~Blusk}
\author{J.~Butt}
\author{E.~Dambasuren}
\author{O.~Dorjkhaidav}
\author{J.~Haynes}
\author{N.~Horwitz}
\author{N.~Menaa}
\author{G.~C.~Moneti}
\author{R.~Mountain}
\author{H.~Muramatsu}
\author{R.~Nandakumar}
\author{R.~Redjimi}
\author{R.~Sia}
\author{T.~Skwarnicki}
\author{S.~Stone}
\author{J.C.~Wang}
\author{Kevin~Zhang}
\affiliation{Syracuse University, Syracuse, New York 13244}
% ...
\author{(CLEO Collaboration)} 
\noaffiliation

\date{\today}

\begin{abstract} 
Using the CLEO detector at the Cornell Electron-positron Storage Ring,
we have measured the scaled momentum spectra, $d\sigma/dx_p$, and the
inclusive production cross sections of the charm mesons \DP, \DZ, \DSP\
and \DSZ\ in \ee\ annihilation at about 10.5~GeV center of mass energy,
excluding the decay products of B mesons. 
The statistical accuracy and momentum resolution are superior to
previous measurements at this energy.
\end{abstract}
\maketitle

\section{Introduction}
We report the measurement of the momentum spectra of charged and neutral
$D$ and \DST\ charm mesons produced at the Cornell Electron-positron
Storage Ring, CESR, in non-resonant \ee\ annihilation at about 10.5~GeV
center of mass energy (CME) and observed with the CLEO detector.  The
\DZ\ and \DP\ spectra each include both directly produced $D$'s, and
$D$'s which are decay products of $D$ excited states.  From them we also
derive the inclusive production cross section for these charm mesons.

While very accurate data on bottom quark production from LEP and SLD
have been published in recent 
years~\cite{Alexander:1995aj,Abe:1999ki,Heister:2001jg,Adeva:1991iw},
the data currently available for studies of charm fragmentation at
10.5~GeV CME~\cite{Argus91,CLEO88}, are quite old and,
by present standards, of poor statistical quality and momentum
resolution.  Our statistical sample is about 80 times larger than the
our previous one~\cite{CLEO88} and our current momentum resolution is
about a factor of 2 better.

The spectra represent measurements of charm quark fragmentation
distributions $D^h_c(x,s)$, \ie, the probability density that a $c$ quark
produces a charm hadron $h$ carrying a fraction $x$ of its
momentum, $\sqrt{s}$ being the ``energy scale'' of the process, the \ee\
CME in 
our case~\cite{biebel,Nason:1993xx}.
Experimental heavy-meson spectra in \ee\ collisions are important for
theoretical and practical reasons: (i) they provide a
component that is not yet calculable in predicting heavy flavor
production in very high energy hadronic collisions,
(ii) they can test
advanced perturbative QCD (PQCD) methods, (iii) they can test the QCD
evolution equations, and (iv) they provide information for best
parametrization of the Monte Carlo simulations on which the analysis of
many high energy experiments partially rely.

Items (i) and (ii) are interconnected.  The calculations of
heavy flavor production cross sections in hadronic collisions (\eg,
at the Tevatron and the LHC) are generally based on the
factorization hypothesis, \ie, a convolution of (a) the parton
distribution function for the colliding hadrons, (b) the perturbative
calculation of the parton-parton cross section and (c) the parton
fragmentation function $D^h_q(x,s)$.  Items (b) and part of (c) (the
parton-shower cascade) can be calculated, in the case of heavy quarks,
using PQCD.  Items (a) and the second phase of (c) (the hadronization
phase) are intrinsically non-perturbative (long distance) processes: as
of now, they must be provided by experiments.  There is an ongoing
theoretical effort to push the potential of PQCD to calculate the
perturbative component of the fragmentation function.  It needs tests
and guidance from the experimental spectra of heavy flavored hadrons
produced in \ee\ annihilation.  De-convolving the calculated PQCD
component from the experimental spectra, one obtains the non-perturbative
component of the fragmentation function.  Unphysical behavior of the
result ({\itshape e.g.}, negative values, extension beyond the
kinematic limit) is
indication that further refinement of the PQCD calculation is needed.
Tests of this kind have been performed up to now on $B$
production in \ee\ annihilation~\cite{Ben-Haim, Cacciari1} and in
hadronic collisions~\cite{Abbott, Acosta-b}, and on charm production
in hadron~\cite{Acosta-c} and $ep$
collisions~\cite{H1-c, ZEUS-c}.  Charm production in \ee\ annihilation
provides a further testing ground of these theoretical
attempts~\cite{Cacciari2}.  The larger value of $\Lambda_{QCD}/m_c$ with
respect to $\Lambda_{QCD}/m_b$ makes these non-perturbative effects more
evident than in bottom hadron production.

Tests of the Altarelli-Parisi evolution
equations~\cite{Altarelli:1977zs,Furmanski:1980cm} have been performed
by our collaboration~\cite{CLEO88} with low sensitivity and over a
relatively small energy interval, comparing the CLEO results with PETRA
results.  The spectra reported in the present paper can be compared with
LEP~\cite{ALEPH99} results providing a test over the 10 to 200~GeV
energy range.

Lacking rigorous calculations of the process of quark and
gluon hadronization, QCD inspired \MC\ simulations have been built: the
Lund String Model~\cite{Artru,Boand,LundSM} and Cluster
Fragmentation~\cite{Marchesini}.  These models have been implemented in
\MC\ programs (JETSET~\cite{jtst74}, UCLA~\cite{Chun},
HERWIG~\cite{Marchesini}).  In each case a number of parameters are
introduced, to be determined by fitting the experimental distributions.
Monte Carlo simulations of quark hadronization are used by experiments
to determine detection efficiencies and to calculate some sources of
backgrounds.  The results presented here include a JETSET
parametrization that produces spectra that agree quite well with the
shapes of all spectra obtained in this analysis.

In all these uses of our results, spectral shapes are most important,
rather than the absolute cross-section values; therefore, shape is the
main focus of our attention. 

In Sec.~\ref{sec:plan} we first list the charm mesons studied in our
analysis along with the decay modes considered and then we describe the
data sample analyzed and outline the procedures used to produce the
spectra.  In Sec.~\ref{sec:MCS} we describe the \MC\ simulations we have
generated and their use. In Sec.~\ref{sec:proc} we give details on
how we extract the signal from the effective mass distributions, and in
Sec.~\ref{sec:deteff} we explain how the detection efficiency is
estimated. Sec.~\ref{check-err} is devoted to discussing the checks we
performed and the evaluation of errors. 
In Sec.~\ref{sec:results} the results, \ie, the charm meson
spectra, are shown in the order given in Sec.~\ref{sec:plan}. 
Our results for the inclusive production cross sections are given
in Sec.~\ref{sec:totcs}.  Our optimization of the JETSET parameters to
reproduce our spectra is described in Sec.~\ref{sec:QQpar}.  In two
appendices we show plots of the detection efficiencies and provide
detailed tables of the measured spectra. 

\section{General Analysis Procedures}\label{sec:plan}

We measure the momentum distributions of \DP, \DZ, \DSP and \DSZ\ using
the following decay modes (charge conjugates are implied throughout this
paper):
\begin{itemize}
\item $D^+$
\subitem $D^+\to K^-\pi^+\pi^+$
\item $D^0$
\subitem $D^0\to K^-\pi^+$
\subitem $D^0\to K^-\pi^+\pi^+\pi^-$
\item $D^{*0}$
\subitem $D^{*0}\to D^0\pi^0\to (K^-\pi^+)\pi^0$
\subitem $D^{*0}\to D^0\pi^0\to (K^-\pi^+\pi^+\pi^-)\pi^0$
\item $D^{*+}$
\subitem $D^{*+}\to D^0\pi^+\to (K^-\pi^+)\pi^+$
\subitem $D^{*+}\to D^0\pi^+\to (K^-\pi^+\pi^+\pi^-)\pi^+$
\end{itemize}

We apply selection criteria to identify events with candidate $D$ and/or
$D^*$ that decay in one of these modes.  We then extract the candidate
$D$ or $D^*$ mass distributions in twenty 0.05 wide bins of the reduced
momentum, $x_p(D)\equiv p/p_{max}$, where $p_{max}$ (approximately
4.95~GeV/c) is the maximum attainable momentum at the relevant beam energy. 

We fit these mass distributions with appropriate signal and background
functions.  The distributions of signal yields vs \xp, corrected for
detection efficiency, give the shape of the \xp\ spectra: 
the main goal of our analysis.  We then divide these spectra by
the integrated luminosity and the appropriate decay branching fractions
to form the differential production cross section $d\sigma/dx_p$ for
each channel.

The use of different decay modes of the same meson provides a check on
possible systematic biases.

The procedures used in the present analyses closely parallel those we
used in measuring $D$ and \DST\ spectra from $B$ decay.~\cite{cleo1997}

\subsection{ Data and Detector}\label{sec-DET-DAT}

The \ee\ annihilation data sample used in this study was taken with
the CLEO II.V detector~\cite{cleo2,silicon} at CESR during 1995--1999.

It consists of 2.9~\fb\ of the ``continuum'' (non-resonant) data sample
at about 10.52~GeV CME (36~MeV below BBbar threshold) and the ``ON4S''
sample, comprising 6.0~\fb\ at 10.58~GeV, the \USSSS\ peak.
Assuming that the shape of the spectrum is the same at these two
energies,\footnote{Comparing our spectra with the corresponding
ones at $\sqrt s=30.4$~GeV~\cite{Petra} we estimated that the fractional
difference between the $D^*$ spectrum at $\sqrt s=10.52$~GeV and the one
at 10.58~GeV is at most 0.075\%, after normalizing one to the other.
Because of this sample merging, our results effectively refer to CME
$\sqrt{s}=10.56~GeV$}  we 
merge the two samples for charm mesons with momenta above the maximum
kinematically allowed in \B\ decay.  For lower momenta we use only the
continuum sample, thus reducing the statistics available in that region.  
All charm hadrons coming from $B$ decays are thereby excluded.

To combine the two parts of the spectra, $x_p <
0.50$ extracted from only the continuum sample, and $x_p > 0.50$
extracted from both the continuum and ``ON4S'' samples, 
using the well known 1/s dependence of the \ee\ annihilation cross
section into a pair of fermions (see Sec.39 of ref.~\cite{PDG}, 2004),
we scale the $x_p < 0.50$ spectra by the factor 
\begin{equation}
  1+\frac{{\cal L}_4}{{\cal L}_0}\frac{s_0}{s_4}= 
  1+\frac{6.0}{2.9}\cdot\frac{(10.52)^2}{(10.58)^2}.\label{lumcorr}
\end{equation}
Here ${\cal L}_0$ and ${\cal L}_4$ are the integrated luminosities
of, respectively, the ``continuum'' and ``On4S'' samples, and $s_0$ and
$s_4$ are the squares of the respective CMEs.  
The statistical sample for $x_p<0.50$ is a factor of three smaller than
that for $x_p>0.50$.

The spectrum so obtained is then divided by the integrated
luminosity, $({\cal L}_0 + {\cal L}_4)$, and by the appropriate decay
branching fraction to obtain $d\sigma/dx_p$ for each channel. 

\subsection{Selection Criteria}\label{sec:selcrit}

We select events using standard CLEO criteria designed to efficiently
select $e^+e^-$ annihilation into hadrons, while rejecting Bhabha 
scattering, $e^+e^- \to \mu^+\mu^-$, and beam-gas interactions.  At least
three tracks are required.  Events with three or four tracks must
also have 65\% of the center-of-mass energy deposited in the calorimeter.
For those with five or more tracks the visible energy, summing
both energy in tracks and neutral energy in the calorimeter,
must exceed 20\% of the center-of-mass energy.

Tracks used to reconstruct a $D$ or $D^*$ are required to be
the result of good tracking fits and to have an angle with respect to the
beam line, $\theta$, such that $|\cos\theta| < 0.91$.   They are also
required to be consistent with originating from the luminous region.
Further, if they have momentum greater than 250 MeV/c, we require that the
impact parameter with respect to the beam line be less than 3~mm, and
that the distance between the point of closest approach to the beam line
and the event vertex be less than 2.5~cm.

We impose particle identification requirements based on specific
ionization (dE/dx) and time of flight measurements for the track.  The
requirement is that the combined \chisq\ probability of the chosen
identification must be greater than 4\%. 

Photon candidate showers detected in the central barrel region
($\vert\cos\theta\vert<0.707$) of the crystal calorimeter are required
to have a minimum energy of 30 MeV.  Those detected in the forward
calorimeters are required to have a minimum energy of 50 MeV.  Photon
candidates are also required to be well separated from the extrapolated
position of all tracks, and the lateral shape of the energy distribution
must be consistent with that expected from an electromagnetic
shower.

Candidate $\pi^0$ mesons are reconstructed from pairs of photon
candidates. At least one of the two must be in the central barrel
region.  To improve the determination of the $\pi^0$ momentum, the two
photon combination is kinematically fitted to the nominal $\pi^0$
mass.  The combination is accepted if this fit has $P(\chisq)\ge10\%$.
The resulting $\pi^0$ 4-momentum is used in $D^{*0}$ reconstruction.

\section{Monte Carlo Simulations}\label{sec:MCS}

Monte Carlo simulations are used to estimate detection efficiencies.
Continuum \ee\ annihilation events are generated using the JETSET 7.3
\cite{SJO} package.  The simulated events are then processed through a
GEANT-based~\cite{GEANT} simulation of the CLEO detector and
reconstructed and analyzed as real data.

The Monte Carlo simulations are also used for other purposes: (i) to
provide a shape for the signal in the candidate $D$ mass distribution
(Sec.\ref{sec:proc}), (ii) to estimate the $D$ and $D^*$ momentum
resolution (Sec.\ref{sec:momres}), and (iii) to perform checks on
the validity of our analysis procedures (Sec.\ref{smoosec}
,\ref{sec:checks}).

We use two kinds of \MCS s.  In the first kind, the ``signal \MC'',
only \eecc\ \ events are generated at the JETSET stage, and an event is
accepted only if the charm meson under study is present.  That meson is
made to decay only in the mode under study.  The corresponding
anti-charm hadron decays generically.  We produce three signal Monte
Carlo's, one for \DP\ and two for \DZ\ for the two decay channels
analyzed.  The $D$'s in these signal Monte Carlo's are the mix of
directly produced $D$'s and $D$'s that are decay products of $D^*$'s and
other excited charm states.  The mix is as generated by the physics
simulation (JETSET).  It follows that each one of these signal Monte
Carlo's act also as signal Monte Carlo for $D^*$'s decaying into that
specific $D$ channel.

In the second kind of simulation, the ``generic \MC'', all possible \ee\
hadronic annihilations are produced according to present
knowledge~\cite{PDG}.

The three signal \MC 's and the generic \MC\ accurately reproduce the
\D\ and \DS\ 
signal shapes observed in data.  Backgrounds in the signal
\MC\ mass distributions are much smaller than those in the generic
\MC, which simulates more accurately the backgrounds in the data.

Both kinds of \MCS\ are used to estimate the detection efficiency.  For
each $D$ or \DS\ meson and its decay chain, we find that the signal
Monte Carlo and generic Monte Carlo-derived efficiencies are
statistically compatible.  This proves that the strong background
reduction in the signal Monte Carlo does not affect the efficiency
estimation or, vice versa, that the large background of the generic
Monte Carlo introduces no appreciable bias in the detection efficiency.

The two statistically independent \MC\ simulations allow internal checks
of our procedures.  We will refer to these as ``generic \MC\ checks''.
In a generic \MC\ check, we analyze 
the generic \MC\ as data, using the procedure to be checked. Then we
correct the reconstructed momentum spectrum using the detection
efficiency obtained from the signal \MC.  Finally we compare this
efficiency-corrected spectrum with the JETSET-generated spectrum that
was the input to the generic \MC.  This comparison consists in
calculating the \chisq\ of the bin-by-bin difference between the
reconstructed and the input spectrum:
\begin{equation}
\chisq=\sum_{i=1}^{n}\left(\frac{R_i-I_i}{\delta R_i}\right)^2,
\end{equation}
where n is the number of bins, $R_i$ and $I_i$ are the values of,
respectively, the reconstructed and input spectra in bin $i$ and $\delta
R_i$ is the statistical error on $R_i$ (the statistical errors on the
input spectra are negligible).  The resulting $\chi^2$ probability, or
confidence level (CL), is the measure of the correctness of the analysis
procedure being checked.
If we normalize the two spectra to each other and recompute
the \chisq, the new CL is a measure of the correctness of our procedure in
so far as the reconstruction of the shape of the spectrum is concerned,
irrespective of normalization.

In a generic \MC\ check, the comparison is with the input spectrum.
It is sensitive to all sources of systematic error on the shape
of the spectra, except for possible errors in physics and detector 
simulation, that are common to signal and generic Monte Carlo.
Hence, insofar as the MC is correct, each check provides 
a comprehensive estimate of all systematic errors associated with the
shape of the spectrum, for the procedure being checked.

\subsection{Momentum Resolution}\label{sec:momres}

\begin{figure}[htb]\center
\includegraphics*[width=2.5in]{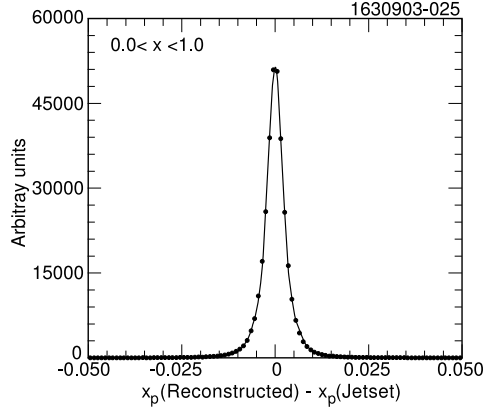}
\vspace{-0.2in}
\caption{\label{fig:x-resol}Resolution in $x_p$ for the \DZ\TO\KM\PIP\
channel. All momenta.}\end{figure}

Comparison with theoretical calculation may involve the moments of the
spectra: $\int_0^1 x^N{d\sigma\over dx}dx$.  In order to minimize
correlations between adjacent \xp\ bins, the \xp\ bin size should be
chosen to be substantially larger than the \xp\ resolution.  It is then
important to know the momentum, and hence the \xp, resolution in our
analysis.  Using the CLEOG \MC\ simulation~\cite{GEANT}, which
reproduces rather accurately our track and shower measurement errors, we
plot the difference between the reconstructed $x_p$ and input $x_p$ (from
JETSET).  Fig.~\ref{fig:x-resol} shows this resolution distribution for
the mode \DZ\TO\KM\PIP\ for all momenta. The full width at half maximum
(FWHM) is 0.008, \ie, 16\% of the bin size (0.050).  The resolution
(FWHM) varies monotonically with momentum, from 4\% of bin size at
$x_p=0.10$ to 18\% for $x_p=0.95$.  For the other channels the
resolution is likewise a small fraction of the bin size.

\section{Candidate mass distribution fitting}\label{sec:proc}

For the \DP\ and \DZ\ analyses we select candidate daughters, add their
four-momenta, and calculate the invariant mass \mdc\ of the charm meson.
Multiple candidates in the same event are accepted. 

In the \DS\ case we obtain the \mdc\ distribution for the \DZ\
associated with the \DS\ by selecting \DS\ candidates with $Q\equiv
\mdsc-\mdc-m_\pi$ in the signal region for \DS\ decay.  Here \mdsc\ is
the invariant mass of the decay products of the candidate $D^*$.  Random
\D-\PI\ associations are subtracted using the \mdc\ distribution for
events in the side bands of the \DS\ signal in the
$Q$ distribution.\footnote{The signal and the side-band regions
are defined as follows.  We fit the ``global'' (\ie\ all momenta) Q
distribution with a Double-Gaussian plus suitable background.  The ratio
SIG2/SIG1 of the widths of the two Gaussians is, in all cases, about
2.2.  We choose the signal region to be MEAN\PM n*SIG2, where n (that
turns out to be about 2 in all channels) is evaluated from the Gaussian
Integral tables, requiring that the whole area of the narrow Gaussian
plus the area within \PM n*SIG2 of the wider Gaussian result in a 98\%
of the Double-Gaussian area.  For the side bands, on each side, we skip
n*SIG2 and then take a region n*SIG2 wide.} 

\begin{figure*}[htb]\center
\includegraphics*[width=6.3in,height=2.5in]{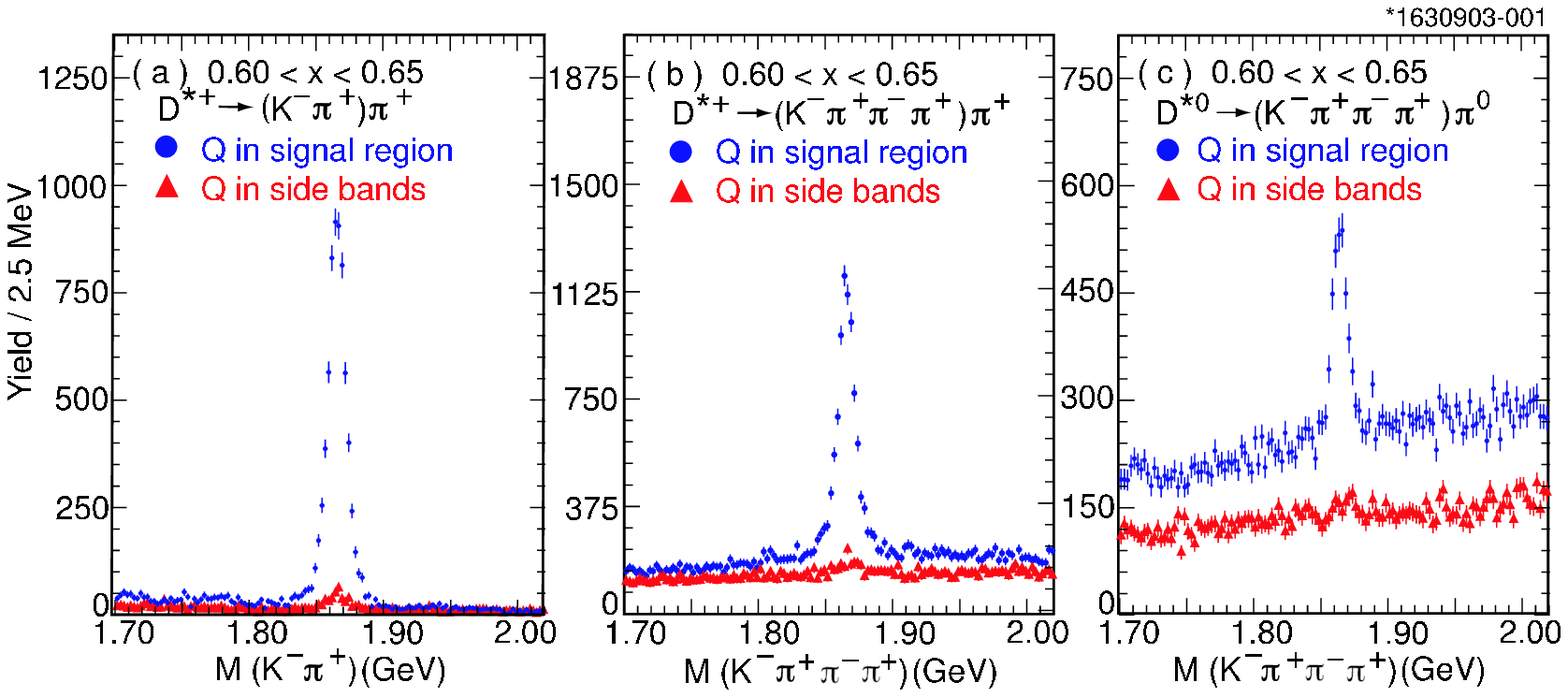}
\vspace{-0.2in}
\caption{\label{fig:ds-4-5}Examples of \mdc\ distribution for two \DSP\
decay channels and one of the \DSZ\ channels analyzed.  (a)
\DSP\TO(\KM\PIP)\PIP.  (b) \DSP\TO(\KM\PIP\PIP\PIM)\PIP.  (c)
\DSZ\TO(\KM\PIP\PIP\PIM)\PIZ.  They show the \mdc\ distribution for 
$Q$ in the \DSP\ signal region and for $Q$ in the \DSP\ side bands.} 
\end{figure*}

Fig.~\ref{fig:ds-4-5} shows examples of the \mdc\ distributions for
three different \DS\ decay modes, for events with $Q$ in the signal 
region and for those 
in the $Q$ side bands.  The residual background after the subtraction
is due to \D\ candidates from random track association. 

The choice of the signal shape used to fit the \mdc\ distribution was
studied and discussed in detail in a previous paper~\cite{cleo1997}.  A
Gaussian function does not give a sufficiently accurate parametrization
of the $D$ signal.  Track measurement errors vary
because of the geometrical orientation of the $D$ decay products
in the detector, because of different momenta of the decay tracks and
overlap with other tracks.  That study concluded that a satisfactory
choice for the $D$ signal shape is a double Gaussian, \ie, the sum
of two Gaussians constrained to have the same mean.  
A different choice of signal fitting function is the signal shape
obtained from the \MCS\ where, for each track, we can identify the
input particle that generated it.  We call the signal mass histograms
thus obtained (one for each momentum bin) the ``TAGMC shape''.
To compare these two choices we repeat a test that was performed in the
previous paper~\cite{cleo1997}, on the \DZ\TO\KM\PIP\ channel, as
follows. 

\begin{figure*}[hbt]\center
\includegraphics*[height=2.5in]{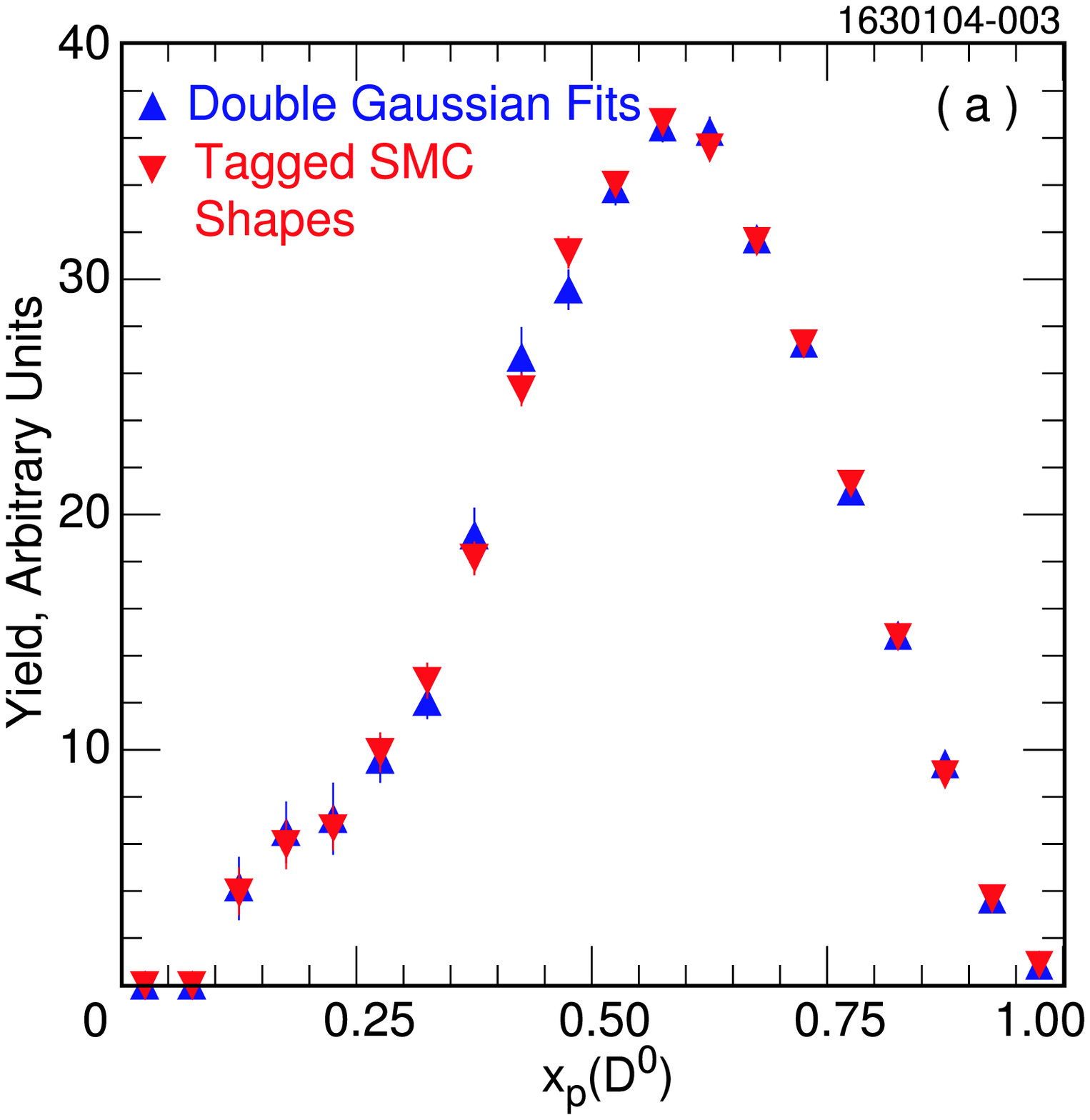}
\includegraphics*[height=2.5in]{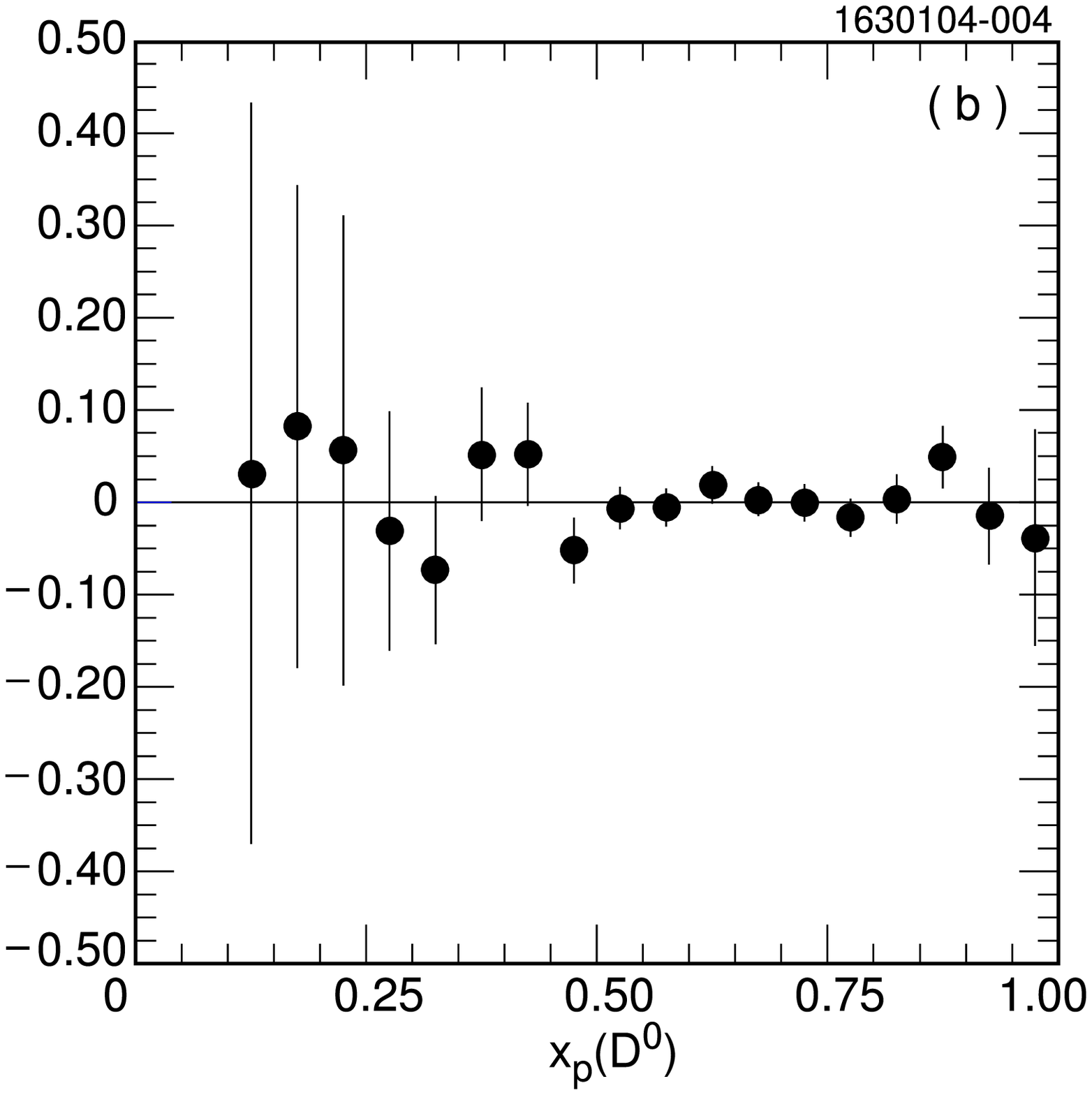}
\vspace{-0.2in}
\caption{\label{fig:tagmc}(a) Overlay of \DZ\ spectra (data) from double
Gaussian and TAGMC shape signal fitting; (b) fractional difference of
the two spectra.}\end{figure*}

We repeat the \DZ\ data analysis, replacing the
double Gaussian with the TAGMC shape.  With this signal shape we obtain
excellent fits, although not superior to the double Gaussian fits.  We
use MINUIT to find the compatibility of the two spectra.  We fit one
using the other as fitting function.  The fitted relative normalization
parameter is $1.016\pm0.007$, and the CL of the fit is 93.8\%.  The
two spectra are compared in Fig.~\ref{fig:tagmc}(a) after normalizing
one to the other.  To find if there is any $x_p$ dependence of the
difference between the spectra obtained by the two methods, we took the
bin-by-bin fractional difference between the two spectra
(Fig.~\ref{fig:tagmc}(b)) and fitted it to a constant, resulting into a
CL=91.0\%, consistent with no difference between the two choices of
signal shape. 
The results obtained using the double Gaussian as signal shape, are
compared with the TAGMC shape to estimate the systematic error on the
total cross sections due to the uncertainty on the signal shape.

The suitability of the double Gaussian as fitting function is also
confirmed by the goodness of the fits: in all the channels, the fit
confidence levels are evenly distributed between 0.0 and 1.0, as they
should be.  A quadratic polynomial is used to fit the combinatoric
background in each of the seven channels.

The fits of the \mdc\ distributions are over the whole 1.70-2.02~GeV
range shown in the figures, except for the $D^+\to K^-\pi^+\pi^+$ case,
where we exclude the 1.96-2.02~GeV ($D^{*+}$) region, and for the
$D^0\to K^-\pi^+$ case, as explained in the next subsection.  The fitted
area of the double Gaussian (or the result of the COUNT procedure
described in Sec.~\ref{subsec:DZ3}, below) is the ``raw'' yield for that
\xp\ bin.

In the next two subsections, we discuss additional backgrounds
in the \mdc\ distribution from the $D^0\to K^-\pi^+$ channel, and describe
an alternative procedure, the COUNT method, to estimate the raw yield
in the $D^0\to K^-\pi^+\pi^-\pi^+$ channel.
 
\subsubsection{The $D^0\to K^-\pi^+$ case}\label{subsec:DZ1}

In the $D^0\to K^-\pi^+$ case (direct or from \DS\ decay) additional
backgrounds must be considered:  \DZ\ decays to \KM\KP,
\PIM\PIP, \KM\RHP\ and \DZB\TO\KP\PIM\  misinterpreted as \KM\PIP.
The shapes of their \mdc\ distributions are obtained from \MC\
simulation. 

The \KM\RHP\ background is very small and contributes only to the
$1.70<M(\KM\PIP)<1.75$~GeV mass region.  This contribution is excluded
by not considering this mass region in the fit.

The background due to \K\PI\ switched identities shows as a
very broad enhancement centered at the signal position.  For $x_p>0.20$,
this enhancement is so broad that it can be easily accommodated by the
quadratic term of the polynomial background function.  For small
\xp, it is narrower, but contributes negligibly.
The amount of this background is fixed to a momentum dependent fraction
determined by \MC\ simulation.

The backgrounds due to \DZ\ decays to \KM\KP, \PIM\PIP\ do not
contribute to the peak, but, if ignored, would result in a very poor fit 
of the background.
Such a fit overestimates the amount of background under the signal and
thus underestimates the amount of signal. The \DZ\TO\KM\KP\ background level
is a parameter to be fitted.  Because of lack of statistics, the amount
of \DZ\TO\PIM\PIP\ background is constrained to a fixed fraction (0.357) of
the \DZ\TO\KM\KP\ background, based on the known relative branching
ratio~\cite{PDG}.  The $\pi\pi$ contribution is very small, and
alternative methods of accounting for it cause negligible changes in
signal yields.

\begin{figure*}[hbt]\center\leavevmode
\includegraphics*[width=6.3in]{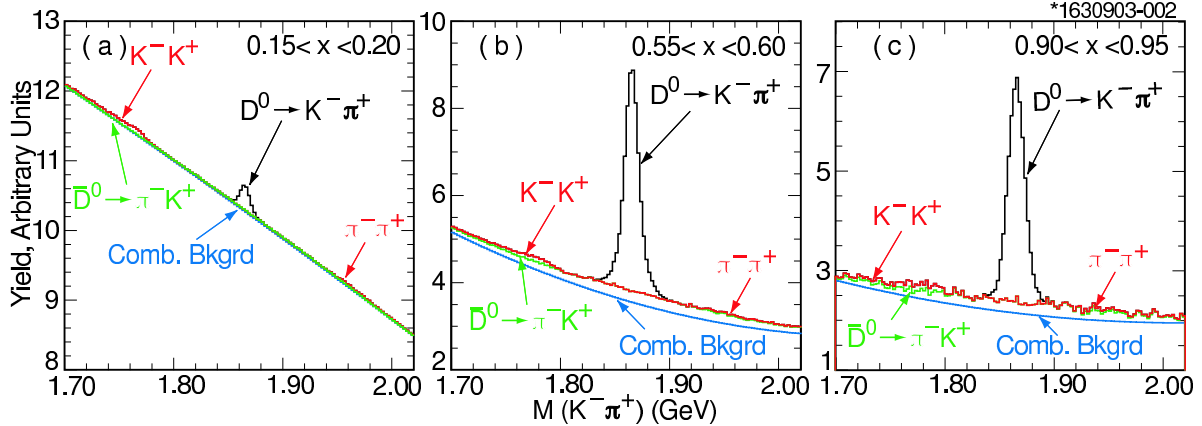}
\vspace{-0.2in}
\caption{\label{fig:dz11}Buildup of the background from its components 
to fit the $M(\KM\PIP)$ distribution.  The solid histogram is
data.  Notice the offset on the yield axis.}\end{figure*}

Fig.~\ref{fig:dz11} shows data in three representative momentum
intervals, demonstrating how the background is built up from the four
contributions.  All four background components are needed to extract the
yield.

\subsubsection{The $D^0\to K^-\pi^+\pi^+\pi^-$ case: the COUNT method}
\label{subsec:DZ3}

In the case of the $D^0\to K^-\pi^+\pi^+\pi^-$ decay, direct or from
\DS\ decay, in addition to using a double Gaussian as fitting function
for the signal, we use a different procedure that leads to results that
are statistically competitive.  In the $D^0\to K^-\pi^+\pi^+\pi^-$ case,
the signal is quite narrow and the background is smooth over a wide
\mdc\ region.  We exclude the signal region and fit the background to a
polynomial.  The signal region is centered on the mean of the double
Gaussian fit and its range is chosen so as to contain the entire signal.
We then count all events in the signal region and subtract the
background obtained from this fit.  The result of this subtraction is
the measured signal yield.  We perform this procedure on data for three
choices of the signal region: 1.810-1.920~GeV, 1.820-1.910~GeV and
1.830-1.900~GeV.

We repeat this procedure on the generic \MC, thus performing the generic
\MC\ check, described in Sec.~\ref{sec:MCS}.  The 1.820-1.910~GeV
exclusion gives the best CL: 28\%.  The narrower exclusion gives the
worst CL: 6\%.  The wider exclusion gives an acceptable CL: 22\%, in
part, because the wider the exclusion region is, the larger the
statistical error becomes.  Based on these results, we choose the data
spectrum obtained with the 1.820-1.910~GeV exclusion as our result.  The
bin-by-bin rms spread of the three data spectra obtained with different
signal region exclusions is taken as the estimate of the systematic error
of this procedure.

We have two valid measurements, one from the COUNT method and the
other from double Gaussian fitting of the signal, both 
performed on the same statistical sample. Hence we take as result the
bin-by-bin arithmetic average of the spectrum obtained by double
Gaussian fitting and the one obtained by the COUNT method with the
optimal choice of the signal region exclusion: $1.820<\mdc<1.910$~GeV.

\subsection{Fit parameter smoothing}\label{smoosec}

The shape parameters of the signal and background functions are expected
to depend smoothly on $x_p$.  By imposing this smoothness of the shape
parameters we suppress, in part, the bin-to-bin (in \xp) statistical
fluctuations in the spectra. This improves the accuracy of the
\textit{shape} of the spectra, particularly at low $x_p$ where
statistics are poor.  This parameter smoothing procedure was used also
in our measurement of charm meson momentum spectra from $B$
decay~\cite{cleo1997}.  In the last paragraph of this subsection we show
the extent of improvement obtained.

The parameters considered are:  the mean of the
double Gaussian (common to the two Gaussians), the width of
the narrower Gaussian, $\sigma_1$, the ratio of the widths of the wider
to the narrower Gaussian, $\sigma_2/\sigma_1$, and the ratio of the area
of the wider Gaussian to the total area, $A_2/A_{tot}$.
We impose this smooth behavior by fitting the \xp\ dependence of each
shape parameter to a polynomial, at most quadratic, in \xp.

We proceed in stages. We start by smoothing the parameter that shows the
least fluctuations and repeat the \mdc\ distribution fitting for all the
\xp\ bins, fixing that parameter to the value given by the smoothing
function.  We do this in sequence for all shape parameters.  If a
parameter does not show appreciable statistical fluctuations, we may
skip smoothing it.  It may take up to five iterations to smooth all
the parameters.

\begin{figure*}[htb] \center\leavevmode
\includegraphics*[width=5.7in]{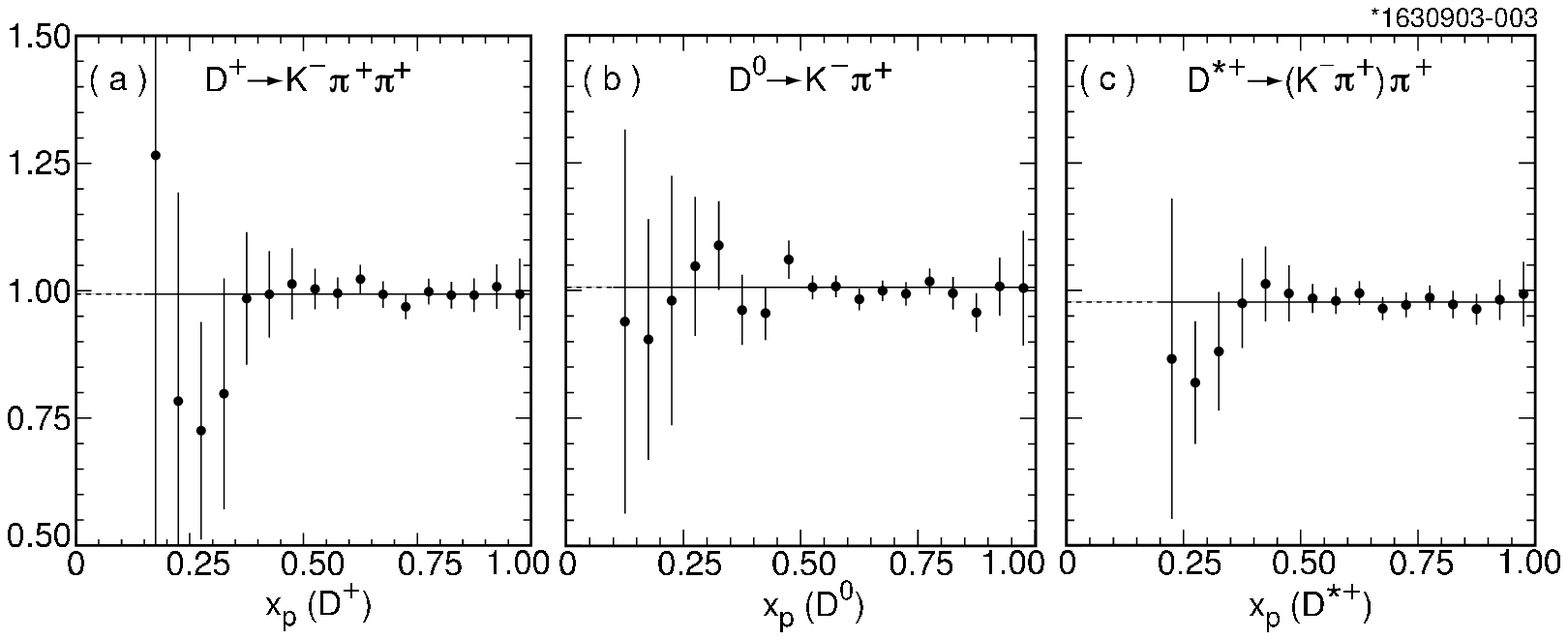}
\vspace{-0.20in}
\caption{\label{fig:smoo-check}Ratios of data spectrum after
double Gaussian shape parameter smoothing to the one obtained without
smoothing: (a) \DP\TO\KM\PIP\PIP, (b) \DZ\TO\KM\PIP, (c)
\DSP\TO(\KM\PIP)\PIP.}
\end{figure*}

At each stage we get a new \xp\ spectrum and check that we have not
introduced any distortion to that spectrum.  The check is performed by
calculating the bin-by-bin ratio of the new spectrum to the original one
where all the parameters were allowed to float (the ``no smoothing''
spectrum).  This ratio should show only random fluctuations around
unity.  If the ratio shows any trend vs \xp, \eg, if a slope and/or a
curvature is needed to describe the \xp\ dependence of the ratio, that
smoothing stage is discarded. Fig.~\ref{fig:smoo-check} shows three
examples of these checks.  When we perform a \chisq\ fit of the ratios
to a constant function (=1), we obtain CL of 94.6\%, 91.0\% and 38.0\%
respectively for the three examples shown.  These are typical for all
the retained smoothing steps.

We perform the smoothing procedure varying the sequence of smoothing
stages.  Each change of sequence leads to a spectrum that is
slightly different from the other ones.  If the CL of the
generic \MC\ check for one of the sequences is considerably higher
than the CL for the other ones, we take that spectrum as our result.  
 
Comparison of spectra derived from different smoothing sequences
provides a measure of the associated systematic error, as explained in
Sec.~\ref{sec:errors}.

We use the generic \MC\ check discussed in Sec.~\ref{sec:MCS} to see if
the smoothing procedure improves the agreement between the reconstructed
and the original spectrum, \ie, the spectrum that is the input to the
\MCS.  In the \DZ\TO\KM\PIP\ case, when there is no smoothing,
the spectrum produced by the analysis fits the original ("true")
spectrum with a $\chi^2=25.1$ for 15 d.o.f., \ie, CL =
5\%.\footnote{Since our aim is to measure the shape of the spectra,
irrespective of normalization, these \chisq\ and related CLs are
calculated after normalizing the reconstructed spectrum to the
original one, thus resulting in an increase of the CLs.} When
smoothing is used, the spectrum produced by the analysis fits the
original spectrum with $\chi^2=7.0$ for 15 d.o.f., \ie, CL = 95\%.
Thus, in this case, parameter smoothing produces a dramatic improvement.
In the case of \DP\TO\KM\PIP\PIP, the CL improves appreciably from 7\%
to 13\%.  In the \DSP\TO(\KM\PIP)\PIP\ case, where the CL is already
93\% without parameter smoothing, there is only a small improvement to a
CL=97\%.  In the \DSZ\TO(\KM\PIP)\PIZ\ case the improvement is from
CL=59\% to CL=75\%.  
As expected, the improvement is strong when the initial set of
parameters show large fluctuations, smaller when the parameters show a
fairly smooth behavior to start with.

\section{Detection Efficiency}\label{sec:deteff}

For each channel we have two independent and statistically-compatible
estimates of the detection efficiency, as explained in
Sec.~\ref{sec:MCS}. We take their weighted average, thus
appreciably reducing the statistical error on the detection efficiency.

The detection efficiency should be a smooth function of \xp.  We use
a second order polynomial to fit the \xp\ dependence of the detection
efficiency averaged over the signal and generic \MC.  Adding a cubic
term does not improve any of the fits.  We call the result of this fit
the ``smoothed efficiency''. In Appendix~\ref{sec:eff.plots}, we show
the detection efficiency dependence on \xp\ for all the mesons and decay
modes analyzed.  In Figs.~\ref{fig:dp6-gmct}, \ref{fig:dspz13-gmct},
\ref{fig:dszz13-gmct}, the detection efficiencies obtained from the
signal and generic \MC 's are plotted, and the curve resulting from the
fit of their average to a polynomial is overlaid.  This procedure
results in a strong reduction of the statistical errors on the detection
efficiency. 

The detection efficiency corrected spectrum is obtained by dividing
the raw signal yield by the smoothed efficiency, bin-by-bin in \xp.

\section{Checks and error estimation}\label{check-err}

\subsection{Two Checks}\label{sec:checks}

\subsubsection{Generic \MC\ checks}
For each procedure used to reconstruct the spectra, we perform a
``generic \MC\ check'', as described in Sec.~\ref{sec:MCS}.  The \CL s
reported below in Table~\ref{tab:gMCcheck}, show the consistency of the
reconstructed spectrum with the original one.  Since our interest is in
the consistency of the shapes of the two spectra, we do the comparison
after normalizing the areas of the the two spectra to each other.  The
normalization differs from unity by at most 2.6\%.  Notice that in the
generic \MC\ checks we can only use the signal \MC\ efficiency, not the
averaged, smoothed efficiency described in the previous section
(Sec.~\ref{sec:deteff}).  

\begin{table}[htb]
\caption{\label{tab:gMCcheck}Confidence levels of the fit of the generic
  \MC\ reconstructed spectrum to its input spectrum for the seven
  decay channels analyzed.}
\begin{center}\begin{ruledtabular}\begin{tabular}{lc|lc|lc}
Decay channel & C.L. & Decay channel & C.L. & Decay channel & C.L. \\ 
$D^+\rightarrow K^-\pi^+\pi^+$ & 18\% & $D^0\rightarrow K^-\pi^+$ & 72\% &
$D^0\rightarrow K^-\pi^+\pi^+\pi^-$ & 56\% \\
$D^{*+}\rightarrow (K^-\pi^+)\pi^+$& 70\% &
$D^{*+}\rightarrow (K^-\pi^+\pi^+\pi^-)\pi^+$ & 37\% &
$D^{*0}\rightarrow (K^-\pi^+)\pi^0$ & 76\% \\ 
 & & $D^{*0}\rightarrow (K^-\pi^+\pi^+\pi^-)\pi^0$ & 99\% &  \\
\end{tabular}\end{ruledtabular}\end{center}\end{table}


\subsubsection{Comparison of spectra from different decay modes}
In the \DZ, \DSP\ and \DSZ\ cases we obtain the respective spectra from
two different \DZ\ decay modes.  We checked that the spectra from the
two different decay modes are statistically compatible.  We calculate
the \chisq\ of the difference, using only the statistical errors.  The
corresponding confidence levels are, respectively, 28\%, 100\% and
0.09\%.  After normalizing one to the other the confidence level become:
85\%, 100\% and 84\%.  This test, however, is not very stringent because
the comparison is dominated by the large statistical errors of the
$D^0\to K^-\pi^+\pi^+\pi^-$ channel.

\subsection{Statistical Errors}\label{sec:staterr}

The statistical errors on the efficiency-corrected yields are obtained by
adding in quadrature the statistical error on the raw yield and the
statistical error on the smoothed efficiency (Sec.~\ref{sec:deteff}).  The
latter is generally considerably smaller than the former.

\subsection{Systematic Errors}\label{sec:errors}

We discuss here systematic errors that could affect the shape of the
differential cross section $d\sigma/dx_p$, although some of them are
found to be independent of \xp.  Additional systematic errors that
affect the normalization of the differential cross section, but not its
shape, will be discussed in Sec.~\ref{sec:totcs} on total cross
sections.  

\subsubsection{Errors found to be independent of \xp\ or negligible.}

We consider the following possible sources of systematic errors: (1) the
choice of signal fitting function, (2) possibly incorrect simulation of
the initial state radiation, (3) effects of swapping between background
curvature and width of the wide Gaussian in \mdc\ distribution fitting,
and (4) effects of low detection efficiency for very low momentum
tracks.

The test, described in Sec.~\ref{sec:proc}, that uses a signal fitting
function other than a double Gaussian, gives us a measure of the
sensitivity of our results to the choice of signal fitting function.
Based on that test, we attribute a systematic error of 1.6\% from the
choice of signal fitting function.  The test shows no momentum
dependence of the difference between the two methods.

We have considered the possibility that inaccurate simulation of initial
state radiation (ISR) may have introduced a systematic error in our
estimate of the detection efficiencies.  We compare the detection
efficiencies discussed in Sec.~\ref{sec:deteff} with those obtained from
\MC\ events where no ISR was produced.  As expected, the latter is
slightly higher than the former, but only by 1.1\%, and its dependence on
\xp\ is negligible.  Since our \MC\ does simulate the initial state
radiation, the uncertainty is only in the accuracy of the simulation.
We thus take half of that, 0.5\%, as contribution to the systematic
error on the cross sections.

Since the momentum dependence of these two uncertainties is found to be
negligible, we take them into account only as errors in the total cross
sections (Sec.~\ref{sec:totcs}). 
 
We considered the possibility of swapping between a background that is
highly curved in the signal region, and the wide component of the double
Gaussian.  The only two channels that show an appreciable background
curvature are \DZ\TO\KM\PIP\PIP\PIM and \DP\TO\KM\PIP\PIP.  In the
first case the full compatibility of the fits with the results of the
COUNT procedure (subsect.~\ref{subsec:DZ3}, CL$>96$\% for both \MC's and
for data), shows that this swapping, if it exists, generates an error
much smaller than the statistical error.  In the \DP\ case we performed
the same test with the same result.

We considered the possibility of errors in the \DSP\ detection
efficiency because of the very rapid decrease in the charged track
detection efficiency for momenta below 120~MeV/c.  The detection
efficiency is practically zero below 70~MeV/c.\footnote{The charged
track detection efficiency has been carefully studied in a series of
CLEO internal documents (unpublished).}  We studied in detail the
momentum distribution of the 
charged \PIPM\ daughter of the \DSPM\ (the ``slow pion'') as a function
of \xp(\DSPM).  Only for $x_p(\DSPM)<0.40$ are there slow
pions with momentum below 120~MeV/c.  From the momentum dependence of
the track detection efficiency and the \DSPM\ isotropic decay
distribution~\cite{D*align}, we can calculate the \DSPM\ detection
efficiency.  The result is consistent with the one resulting from our 
generic and signal \MCS\ within their statistical errors. 

Since we find the errors from these last two sources to be negligible,
we disregard them.

\subsubsection{Errors that affect the spectra shapes}\label{spreads}

The different sequences of parameter smoothing stages (described in
Sec.~\ref{smoosec}) lead to slightly different resulting spectra.  We
calculate the root-mean-square (rms) spreads of the yields for each \xp\
bin over the spectra from different sequences. Since these rms spreads
fluctuate statistically from bin to bin, as expected, we average them
over groups of three bins.  We take these rms spreads as systematic
errors on the yields. 

As stated in Sec.~\ref{sec:MCS}, we have both generic and signal \MC\
samples of events, and to the extent that our \MC\ correctly simulates
data and detector, we can perform a test which give comprehensive
information on all systematic errors associated with our analysis
procedures.
We take the bin-by-bin difference between the generic \MC\ 
reconstructed spectrum and the input spectrum, and divide
this, bin-by-bin, by the input spectrum, resulting in the distribution
of the fractional difference vs \xp.  The weighted average, over the
entire \xp\ range, of the absolute values of these fractional
differences (where the weights are the inverse square errors on the
differences) can be considered as an estimate of the systematic error.
It varies from 0.6\% for the \DZ\TO\KM\PIP\ channel to 1.4\% for the
\DP\TO\KM\PIP\PIP\ channel.  The distributions of the fractional
differences show negligible dependence on \xp, meaning that this
estimated systematic error does not seem to affect the shape of the
spectra.  Nevertheless we include these average differences as a
component of the systematic error on the measured yields.  In principle,
this estimate of the systematic error takes into account also the ``rms
spreads'' discussed in the previous paragraph.  We decided, however, to
be conservative, and have combined them in quadrature to obtain the
total systematic error.  Even with possible overestimate, generally the
systematic error makes the total error larger than the statistical error
by only 10\% to 30\%.

\subsection{Total errors}

The statistical errors and the two systematic errors affecting the
spectra shapes are listed, channel by channel, in
Table~\ref{tab:DPspectrum} - \ref{tab:DSZZ3spectrum} in
Appendix~\ref{sec:tables}.  These three errors are combined in
quadrature to give total errors relevant to the shape of our spectra.

\vspace{-0.2in}\section{Results on the shape of the Spectra}
\label{sec:results}

\vspace{-0.2in}\subsection{The Final or Combined Spectrum.}
\label{sec:combine}

\vspace{-0.1in}
For each $D$ or \DS\ meson and its decay chain, we obtain the spectrum
fitting the signal with a double Gaussian after smoothing the $x_p$
dependence of the Gaussian parameters, as described in Sec.~\ref{smoosec}.
When we also employ the COUNT method, as explained in
Sec.~\ref{subsec:DZ3}, the spectrum that we report is the average of the
spectrum obtained by fitting a double Gaussian and that obtained with
the COUNT method. 
Details specific to each channel, are given in the sections showing the
respective spectra.

The spectra shown in the following are differential, inclusive
production cross sections, $d\sigma(\ee\to D^{(*)} X)/dx_p$ at
$\sqrt s=10.58$~GeV  fully
corrected for detection efficiency and decay branching ratios.
We use the following decay branching ratios:
\BR(\DZ\TO\KM\PIP)=(3.82\PM 0.09)\%, 
\BR(\DZ\TO\KM\PIP\PIM\PIP)=(7.49\PM 0.31)\%,
\BR(\DP\TO\KM\PIP\PIP)=(9.0\PM 0.6)\%, 
\BR(\DSP\TO\DZ\PIP)=(67.6\PM 0.5)\%,
\BR(\DSZ\TO\DZ\PIZ)=(61.9\PM 2.9)\%.  
They affect only the normalization, not the shape, of the spectra.
Uncertainties in the branching ratios will be reflected in the
systematic errors on the total cross sections, Sec.~\ref{sec:totcs}.

\vfill
\newpage
\subsection{\DP\ Spectrum}

\begin{figure*}[htb]\center
\includegraphics*[width=6.3in,height=2.1in]{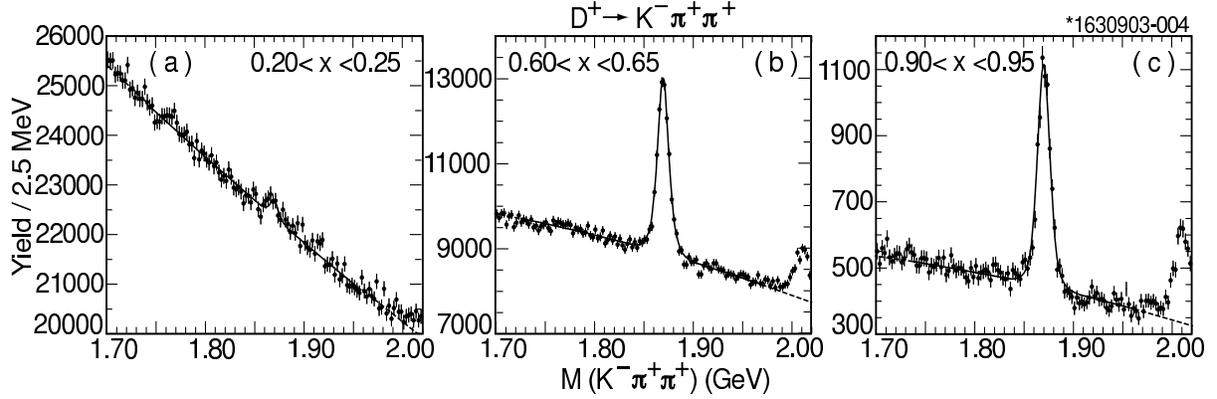}
\caption{\label{fig:dp6-fit}Three examples of $M(\KM\PIP\PIP)$
distribution fits. Notice the large vertical scale offsets.}\end{figure*}

\begin{figure*}[htb]\center
\includegraphics*[width=3.1in]{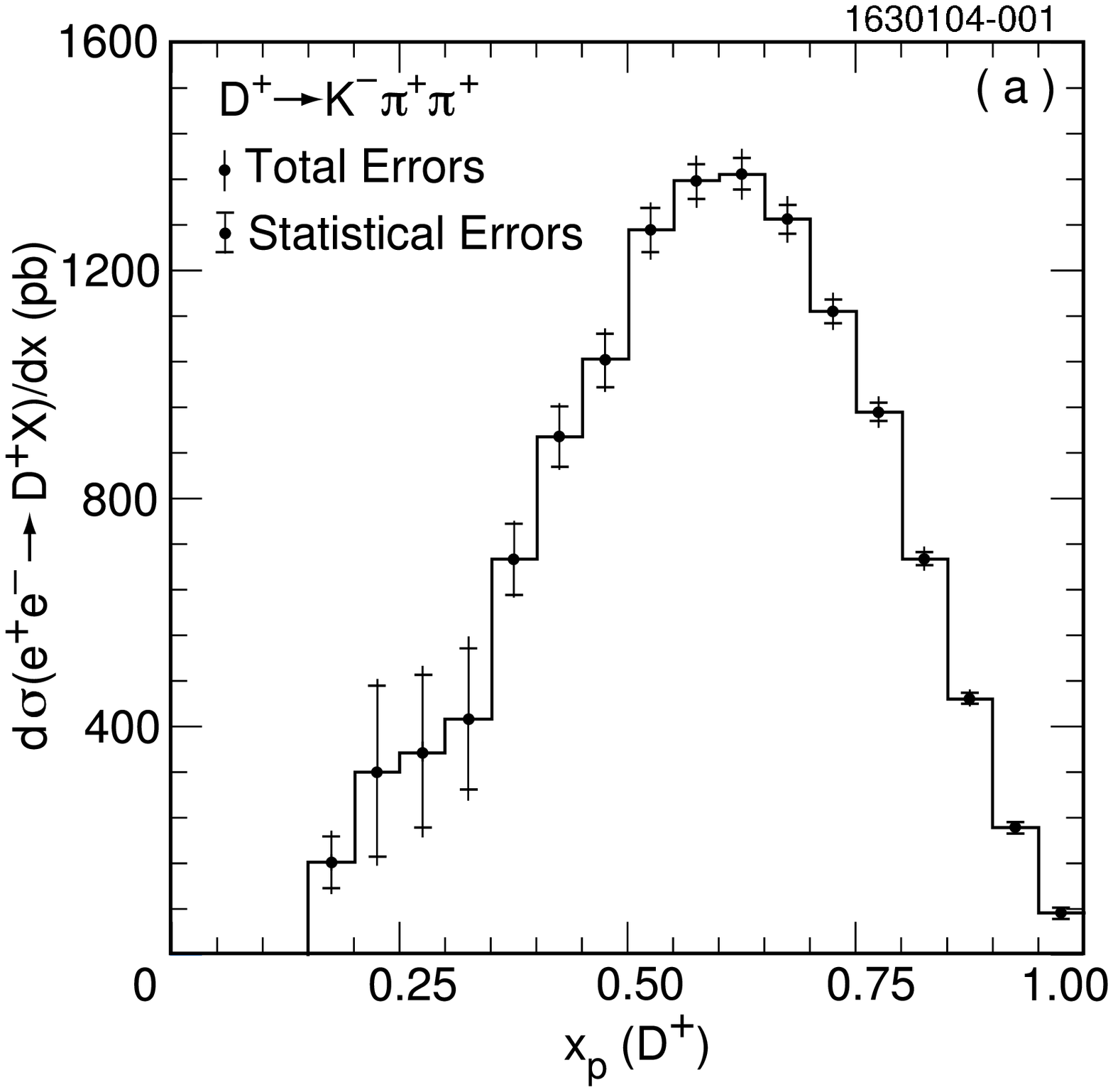}
\includegraphics*[width=3.1in]{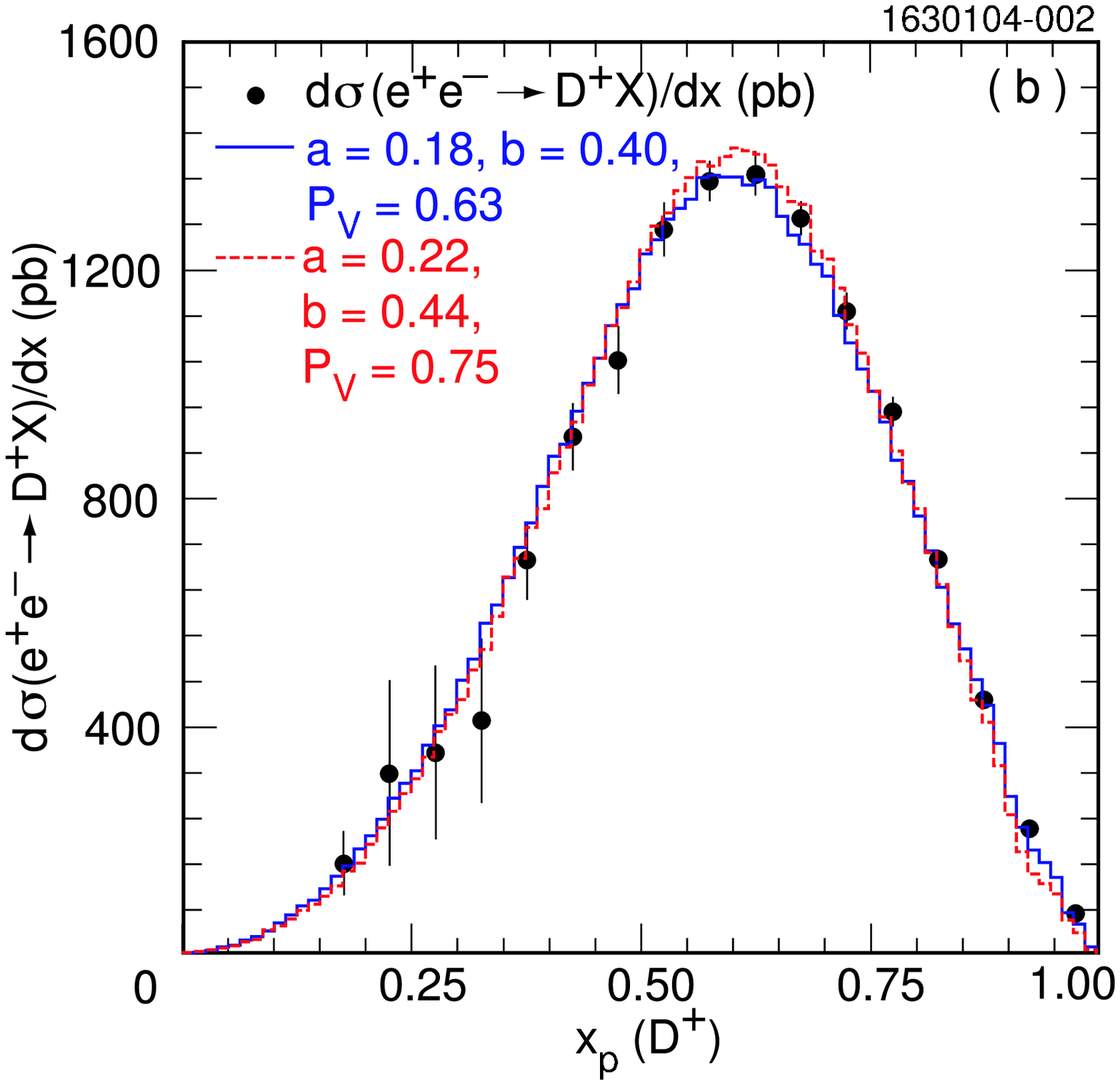}
\caption{\label{fig:DPspectrum}Differential cross section
$d\sigma(e^+e^-\to D^+ X)/dx_p$ in pb from the \DP\TO\KM\PIP\PIP\ decay
mode. (a) shows explicitly the total and statistical errors. (b) the
same spectrum overlaid with the JETSET spectra generated with two
different sets of parameters (Sec.~\ref{sec:QQpar}).}
\end{figure*}

Fig.~\ref{fig:dp6-fit} shows examples of fits to the \mdc\ distributions
in three representative \xp\ bins, using fully smoothed parameters.  Our
result is shown in Fig.~\ref{fig:DPspectrum} and tabulated in
App.~\ref{sec:tables}, Table~\ref{tab:DPspectrum}.  The spectrum shown
is obtained after 
smoothing the \xp\ dependence of the double Gaussian shape parameters
(see Sec.~\ref{smoosec}) using the sequence that gives the best CL in
the generic \MC\ check (Sec.~\ref{sec:checks}).

\subsection{\DZ\ Spectrum}

\subsubsection{\DZ Spectrum from \DZ\TO\KM\PIP }\label{sec:DZ1spect}

\begin{figure*}[htb]\center\leavevmode
\includegraphics*[width=6.3in,height=2.1in]{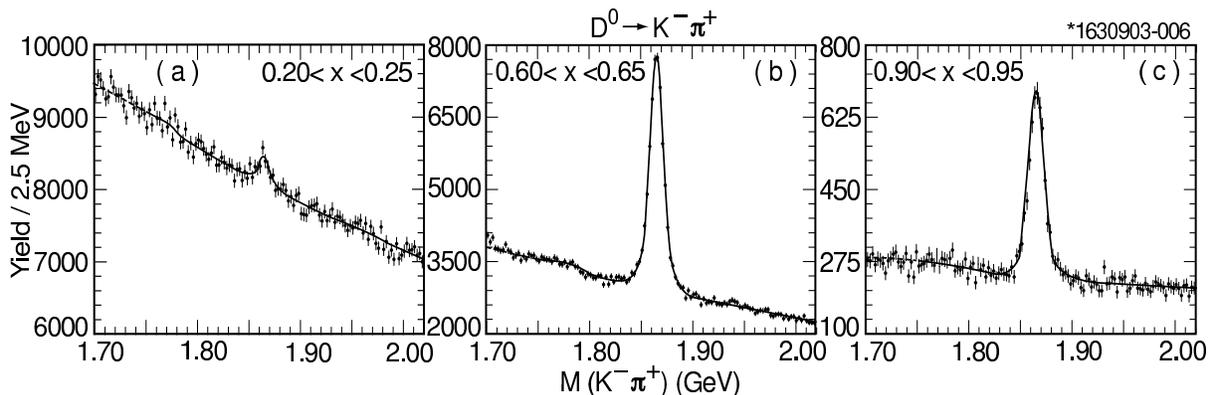}
\vspace{-0.25in}
\caption{\label{fig:dz1-fit}Three examples of $M(\KM\PIP)$ distribution
fits. Notice the large y offsets.}\end{figure*}

Fig.~\ref{fig:dz1-fit} shows examples of fits to the \mdc\ distributions
in three representative \xp\ bins, using fully smoothed parameters.

\begin{figure}[hbt]\center\vspace{0.15in}
\includegraphics*[width=3.2in]{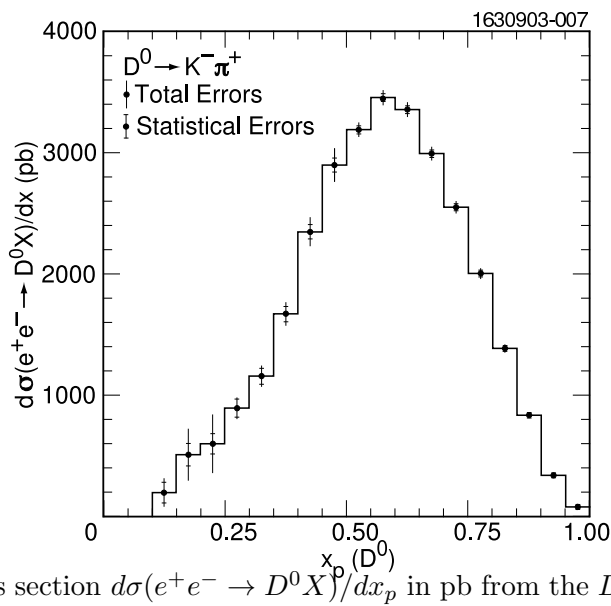}
\vspace{-0.3in}
\caption{\label{fig:DZ1-spectrum}Differential cross section
$d\sigma(e^+e^-\to D^0 X)/dx_p$ in pb from the $D^0\rightarrow K^-\pi^+$
decay mode.}\end{figure}

The \DZ\ inclusive, differential production cross section obtained from
this decay mode is shown in Fig.~\ref{fig:DZ1-spectrum} and in
App.~\ref{sec:tables}, Table~\ref{tab:DZ1spectrum}.  It is obtained
after smoothing the \xp\ 
dependence of the double Gaussian shape parameters (see
Sec.~\ref{smoosec}) using the sequence that gives the best CL in the
generic \MC\ check (Sec.~\ref{sec:checks}).


\subsubsection{\DZ\ Spectrum from \DZ\TO\KM\PIP\PIP\PIM}\label{sec:DZ3spect}

\begin{figure*}[htb]\center\leavevmode
\includegraphics*[width=6.3in,height=2.1in]{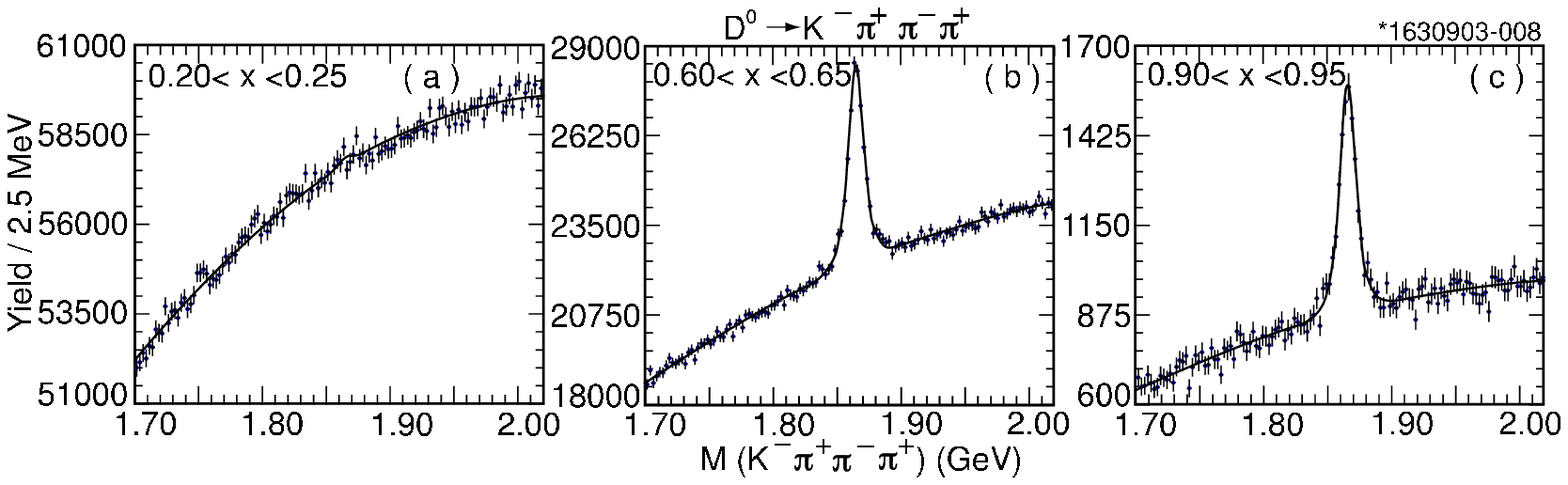}
\vspace{-0.2in}
\caption{\label{fig:dz3-fit}Three examples of $M(\KM\PIP\PIP\PIM)$
distribution fits.  Notice the large y offsets.}\end{figure*}

Fig.~\ref{fig:dz3-fit} shows examples of fits to the \mdc\ distributions
in three representative \xp\ bins, with no parameter smoothing.  Because
of the large statistical errors, we find the Gaussian parameter
smoothing procedure to be unreliable.  However, as discussed in
Sec.~\ref{subsec:DZ3}, for this mode we use also the COUNT method with
three different widths of the excluded signal region in order to get
part of the systematic error on this procedure.

\begin{figure}[hbt]\center
\includegraphics*[width=3.5in]{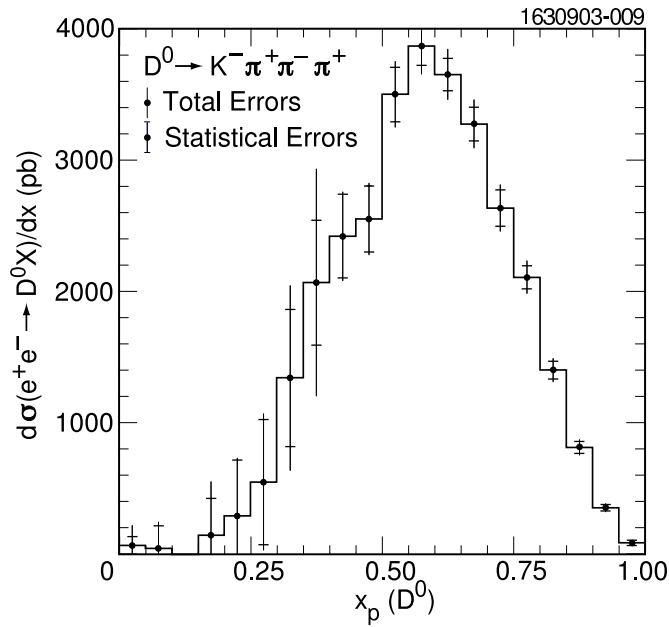}
\vspace{-0.15in}
\caption{\label{fig:DZ3-spectrum}Differential cross section
$d\sigma(e^+e^-\to D^0 X)/dx_p$ in pb from the $D^0\rightarrow
K^-\pi^+\pi^+\pi^-$ decay mode.}
\end{figure}

The \DZ\ inclusive, differential production cross section obtained from
this decay mode is shown in Fig.~\ref{fig:DZ3-spectrum}
and tabulated in App.~\ref{sec:tables}, Table~\ref{tab:DZ3spectrum}.    
It is the arithmetic average of the one obtained by double Gaussian fits
(without any Gaussian parameter smoothing) and the one produced with the
COUNT procedure, excluding from the background fit the 1.820-1.910~GeV
region.  For the final statistical errors we take the average of the 
statistical errors associated with the two methods. 

\subsubsection{The Average \DZ\ Spectrum}

\begin{figure}[htb]\center\vspace{-0.15in}
\includegraphics*[width=3.2in]{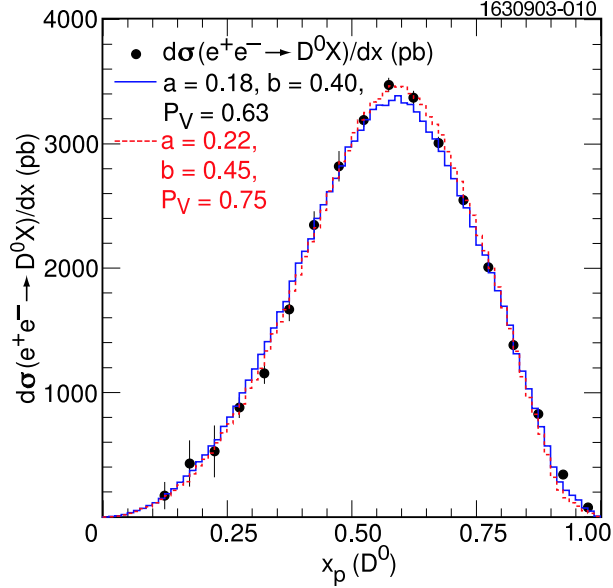}
\vspace{-0.1in}
\caption{\label{fig:DZaver}Differential cross section $d\sigma(e^+e^-\to
D^0X)/dx_p$, weighted average of the spectra from the \DZ\TO\KM\PIP\ and
\DZ\TO\KM\PIP\PIP\PIM\ decay modes, overlaid with the JETSET spectra
generated with two different sets of parameters (Sec.~\ref{sec:QQpar}).}
\end{figure}

The weighted average of the spectra obtained from the two $D^0$ decay
modes analyzed is shown in Fig.~\ref{fig:DZaver} and tabulated in
App.~\ref{sec:tables}, Table~\ref{tab:aver-sp}.  The two JETSET generated
spectra are explained in Sec.~\ref{sec:QQpar}.

\vspace{-0.3in}\subsection{The \DSP\ Spectrum}\vspace{-0.1in}

In Sec.~\ref{sec:proc} we described our procedure for selecting \DSP\
candidates.  The difference between the two \mdc\ distributions shown in
Fig.~\ref{fig:ds-4-5} eliminates random \DZ\PIP\ associations.

\vspace{-0.2in}
\subsubsection{\DSP\ Spectrum from \DSP\TO\DZ\PIP\TO(\KM\PIP)\PIP}
\label{sec:DSPZ1}
\vspace{-0.1in}

\begin{figure*}[htb]\center\vspace{-0.05in}
\includegraphics*[width=6.3in,height=2.1in]{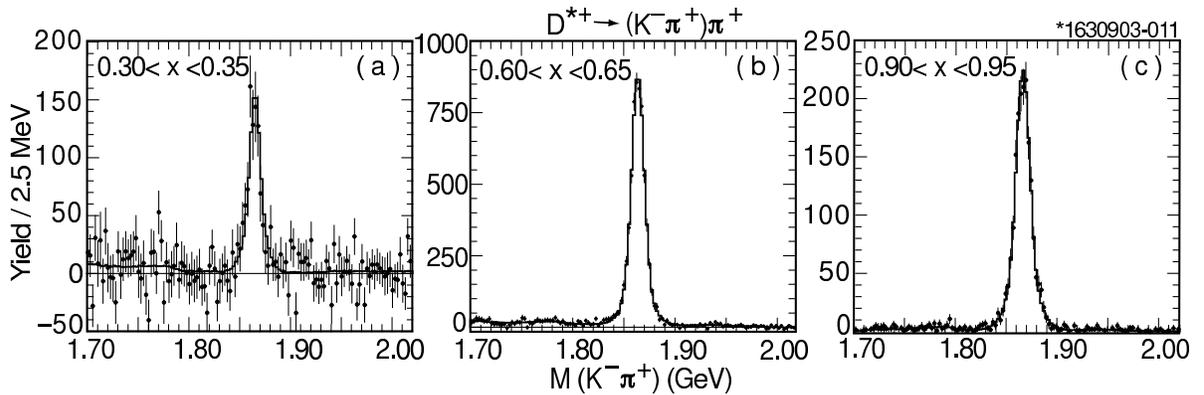}
\vspace{-0.2in}
\caption{\label{fig:dspz1-fit}Three examples of fits of the subtracted
$M(\KM\PIP)$ distributions for $D^{*+}\rightarrow D^0\pi^+\rightarrow
(K^-\pi^+)\pi^+$ candidates.}\end{figure*}

The subtracted \mdc\ distribution (Fig.~\ref{fig:dspz1-fit}) shows the
additional backgrounds present in this \DZ\ decay mode.  They have been
handled as described in Sec.~\ref{subsec:DZ1}.

\begin{figure}[hbt]\center
\includegraphics*[width=3.5in]{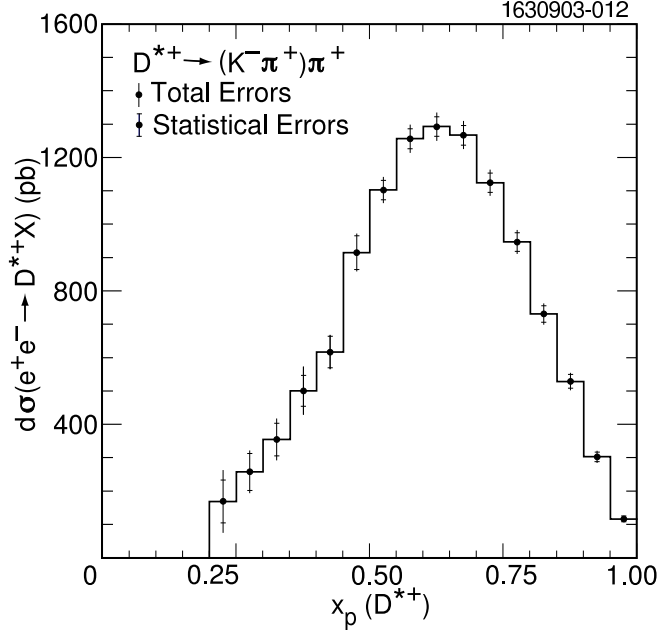}
\vspace{-0.2in}
\caption{\label{fig:DSPZ1}$d\sigma(\ee\rightarrow D^{*+} X)/dx_p$, from
the $D^{*+}\rightarrow D^0\pi^+\rightarrow (K^-\pi^+)\pi^+$ decay mode.}
\end{figure}

The spectrum is shown in Fig.~\ref{fig:DSPZ1} and tabulated in
App.~\ref{sec:tables}, Table~\ref{tab:DSPZ1spectrum}.  It is the one
obtained after smoothing the \xp\ dependence of the double Gaussian
shape parameters (see Sec.~\ref{smoosec}) using the sequence that gave
the best CL in the generic MC check (Sec.~\ref{sec:checks}). 

\subsubsection{\DSP\ Spectrum from \DSP\TO\DZ\PIP\TO(\KM\PIP\PIP\PIM)\PIP}

\begin{figure*}[htb]\vspace{-0.05in}\center
\includegraphics*[width=6.3in,height=2.1in]{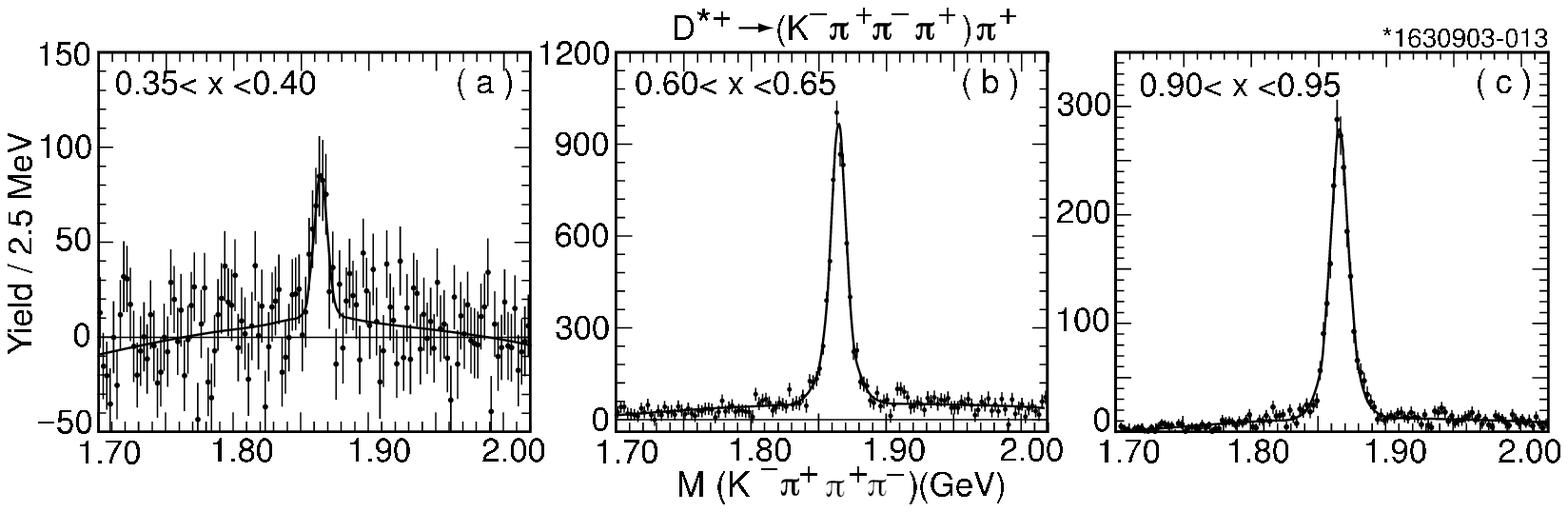}
\vspace{-0.3in}\caption{\label{fig:dspz3-fit}Three examples of fits of
the subtracted $M(\KM\PIP\PIM\PIP)$ distributions for $D^{*+}\rightarrow
D^0\pi^+\rightarrow (K^-\pi^+\pi^-\pi^+)\pi^+$ candidates.}\end{figure*}

Just as in the case of $D^0\rightarrow K^-\pi^+\pi^+\pi^-$, taking
advantage of the narrowness of the signal over a background that is smooth and
well determined over a large region, we use the COUNT procedure
described in Sec.~\ref{subsec:DZ3} with the signal region exclusion as
optimized in that analysis (1.820-1.910~GeV).  The $Q$ selection reduces
drastically the background with respect the \DZ\ case, and we obtain
good double Gaussian fits of the signal as shown, for three
representative \xp\ bins, in Fig.~\ref{fig:dspz3-fit}.

\begin{figure}[bht]\center
\includegraphics*[width=3in]{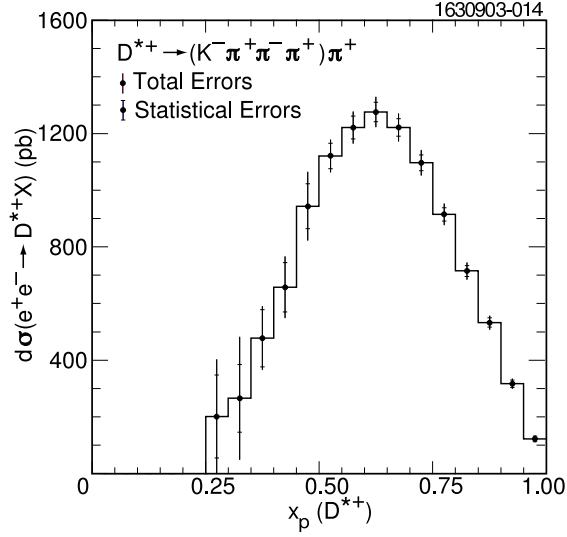}
\vspace{-0.15in}
\caption{\label{fig:DSPZ3} $d\sigma(\ee\rightarrow D^{*+} X)/dx_p$ from
the $D^{*+}\rightarrow D^0\pi^+\rightarrow (K^-\pi^+\pi^-\pi^+)\pi^+$
decay mode.}\end{figure}

The spectrum is shown in Fig.~\ref{fig:DSPZ3} and tabulated in
App.~\ref{sec:tables}, Table~\ref{tab:DSPZ3spectrum}.  It is the
arithmetic average of the one 
obtained by double Gaussian fit, after full smoothing of the \xp\
dependence of the double Gaussian shape parameters (see Sec.~\ref{smoosec}), 
and the one produced with the COUNT procedure, excluding from
the background fit the 1.820-1.910~GeV region.

\subsubsection{The Average \DSP\ Spectrum}

\begin{figure}[htb]\center\vspace{-0.1in}
\includegraphics*[width=3in]{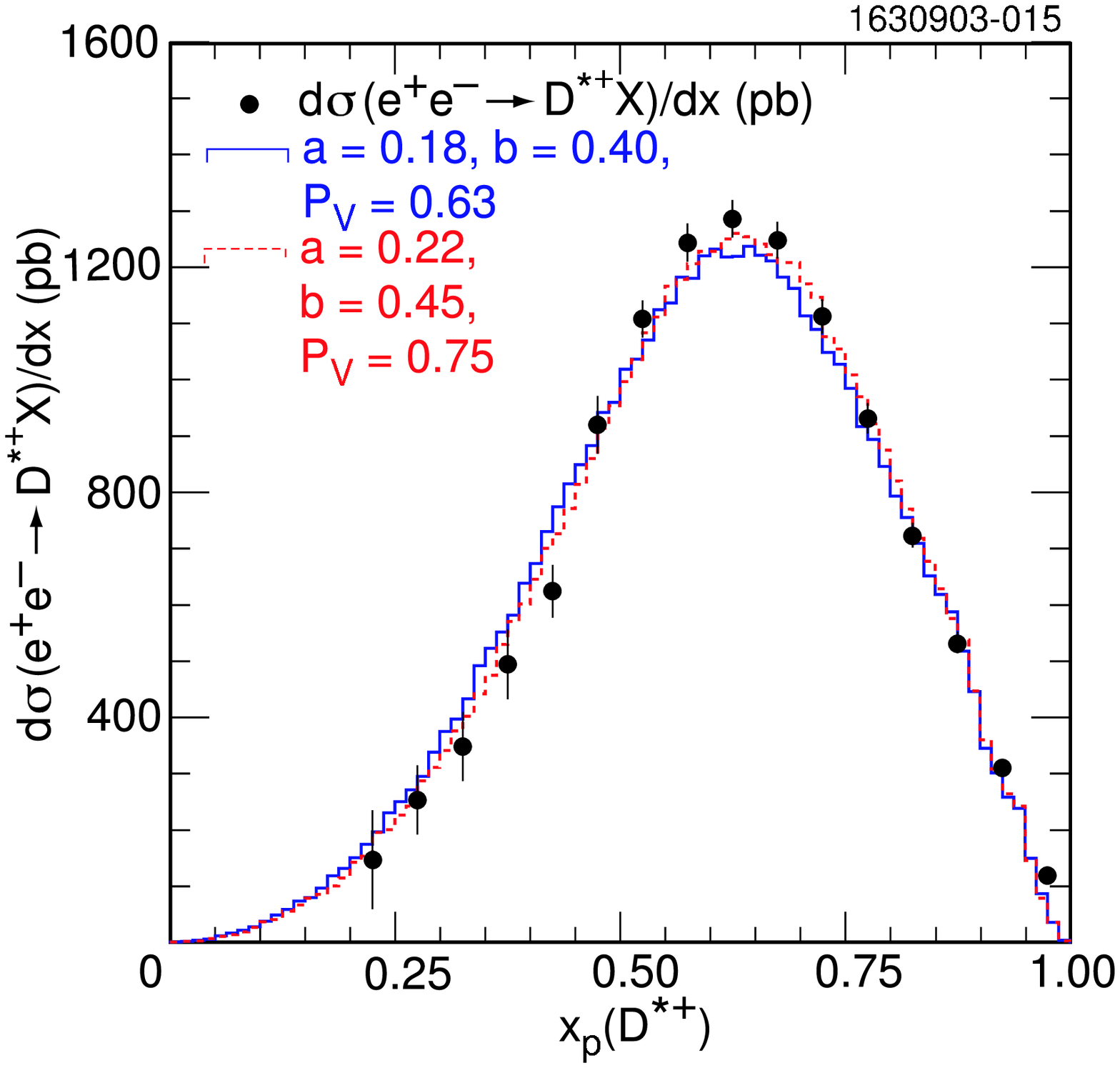}
\vspace{-0.2in}
\caption{\label{fig:DSPaver}Differential cross section
$d\sigma(e^+e^-\to D^{*+}X)/dx_p$, weighted average of
\DSP\TO\DZ\PIP\TO(\KM\PIP)\PIP and
\DSP\TO\DZ\PIP\TO(\KM\PIP\PIP\PIM)\PIP spectra, overlaid with the JETSET
spectra generated with two sets of parameters (Sec.~\ref{sec:QQpar}).}
\end{figure}

The weighted average of the spectra obtained from the two decay modes
analyzed is shown in Fig.~\ref{fig:DSPaver} and tabulated in
App.~\ref{sec:tables}, Table~\ref{tab:aver-sp}.  The two JETSET
generated spectra are explained in Sec.~\ref{sec:QQpar}.

\subsection{\DSZ\ Spectrum}

To suppress random \DZ\PIZ\ associations, we use the subtraction
procedure already used for the \DSP\ cases and illustrated in
Fig.~\ref{fig:ds-4-5}. 
 
\subsubsection{\DSZ\ Spectrum from \DSZ\TO\DZ\PIZ\TO(\KM\PIP)\PIZ}

\begin{figure*}[hbt]\vspace{-0.25in}
\includegraphics*[width=6.3in,height=2.1in]{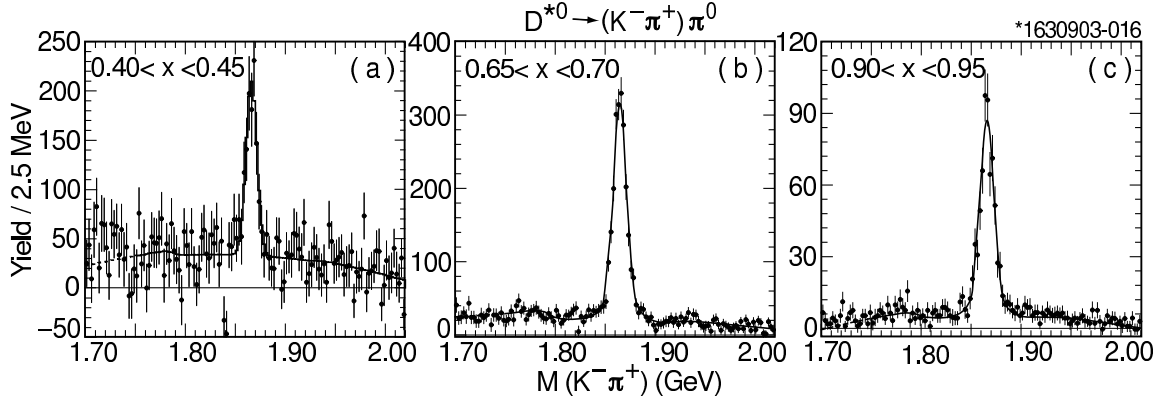}
\vspace{-0.15in}
\caption{\label{fig:dszz1-fit}Three examples of fits of the $M(\KM\PIP)$
distributions for $D^{*0}\rightarrow D^0\pi^0\rightarrow
(K^-\pi^+)\pi^0$ candidates.}
\end{figure*}

Fig.~\ref{fig:dszz1-fit} shows three examples of fits of the subtracted
\mdc\ distribution for this channel. Here too we add to the fitting
functions the backgrounds described in Sec.~\ref{subsec:DZ1}.

\begin{figure}[hbt]\vspace{-0.1in}
\includegraphics*[width=3.5in]{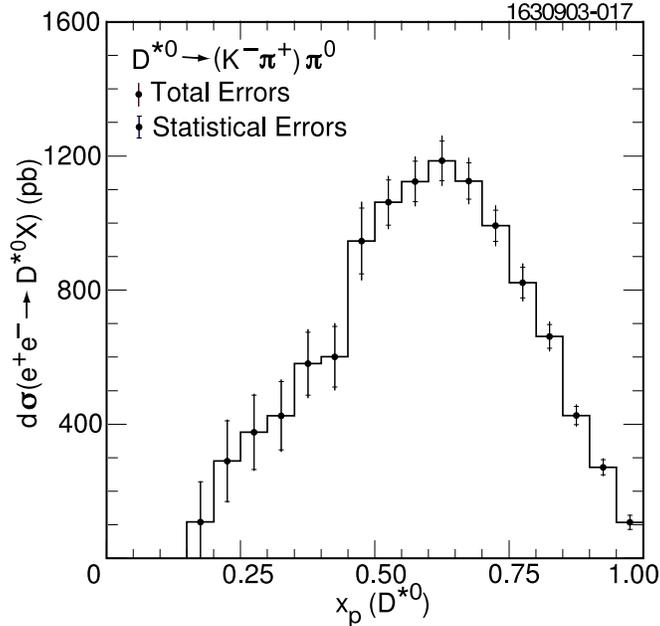}
\vspace{-0.25in}
\caption{\label{fig:DSZZ1} $d\sigma(\ee\rightarrow D^{*0} X)/dx_p$, from
the $D^{*0}\rightarrow D^0\pi^0\rightarrow (K^-\pi^+)\pi^0$ decay mode.}
\end{figure}

The differential cross section is shown in Fig.~\ref{fig:DSZZ1} and
tabulated in in App.~\ref{sec:tables}, Table~\ref{tab:DSZZ1spectrum}.
Among the different 
stage sequences in smoothing the Gaussian parameters (see
Sec.~\ref{smoosec}) we choose the one that gives the best CL in the
generic MC check (Sec.~\ref{sec:checks}).

\subsubsection{\DSZ\ Spectrum from \DSZ\TO\DZ\PIZ\TO(\KM\PIP\PIP\PIM)\PIZ}

\begin{figure*}[htb]\center
\includegraphics*[width=6.5in,height=2.3in]{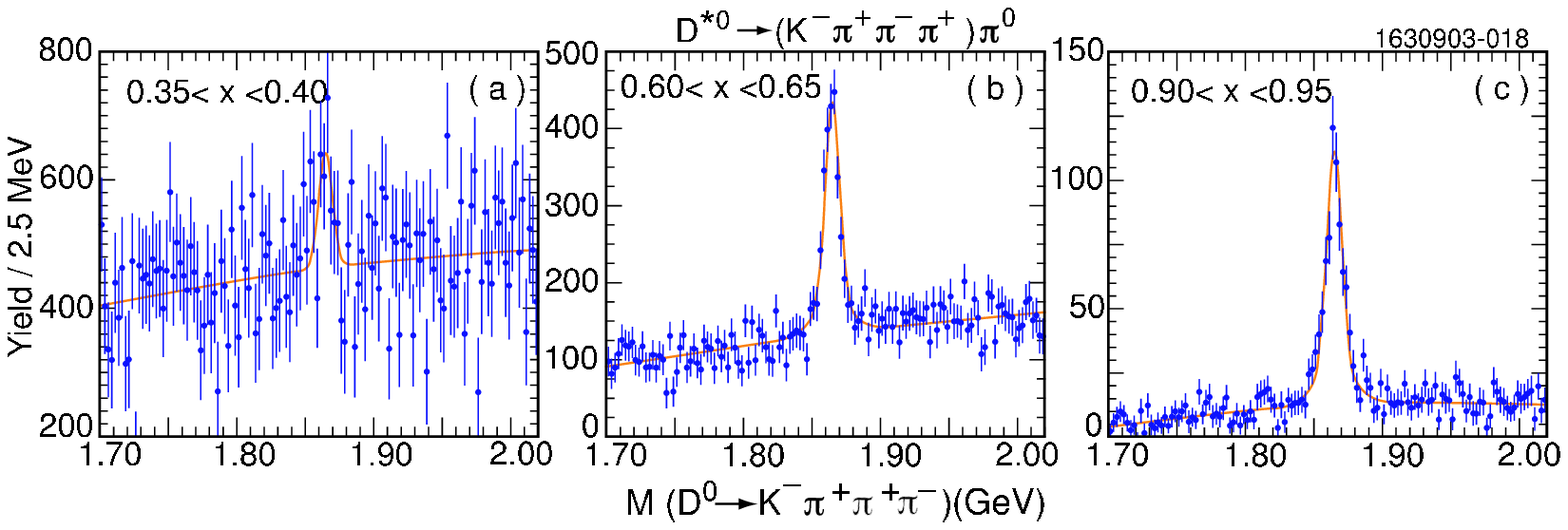}
\vspace{-0.3in}
\caption{\label{fig:dszz3-fit}Three examples of fits of the subtracted
$M(\KM\PIP\PIM\PIP)$ distributions for $D^{*0}\rightarrow
D^0\pi^0\rightarrow (K^-\pi^+\pi^-\pi^+)\pi^0$ candidates.}\end{figure*}

Fig.~\ref{fig:dszz3-fit} shows, for three representative \xp\ bins, the
fits of the subtracted \mdc\ distribution, using a double Gaussian and a
polynomial background. 

Because of the smaller decay branching ratio and the smaller detection
efficiency, due to the presence of a \PIZ, the statistical errors are
quite large, especially for $\xp<0.50$, where we can use only the
continuum events.  We have used both the COUNT procedure and the double
Gaussian signal fitting (without parameter smoothing) to get the \DSZ\
yield.

\begin{figure}[hbt]\center\vspace{-0.1in}
\includegraphics*[width=3.5in]{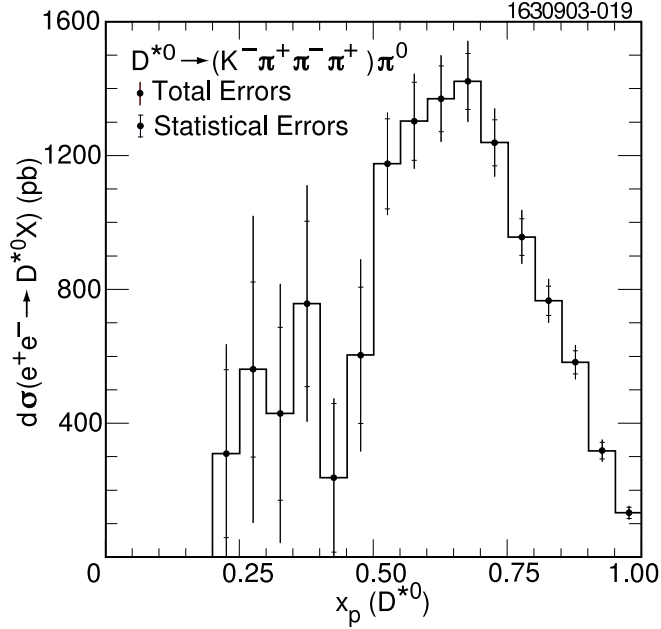}
\vspace{-0.15in}
\caption{\label{fig:DSZZ3} $d\sigma(\ee\rightarrow D^{*0} X)/dx_p$, from
the $D^{*0}\rightarrow D^0\pi^0\rightarrow (K^-\pi^+\pi^-\pi^+)\pi^0$
decay mode.}
\end{figure}

The spectrum is shown in Fig.~\ref{fig:DSZZ3} and tabulated in
App.~\ref{sec:tables}, Table~\ref{tab:DSZZ3spectrum}.  It is the
arithmetic average of 
that obtained by fitting the signal with the double Gaussian (smoothed
parameters) and the one obtained by the COUNT method using the signal
region exclusion optimized in that analysis (1.820-1.910~GeV).

\subsubsection{The Average \DSZ\ Spectrum}

\begin{figure}[hbt]
\includegraphics*[width=3.5in]{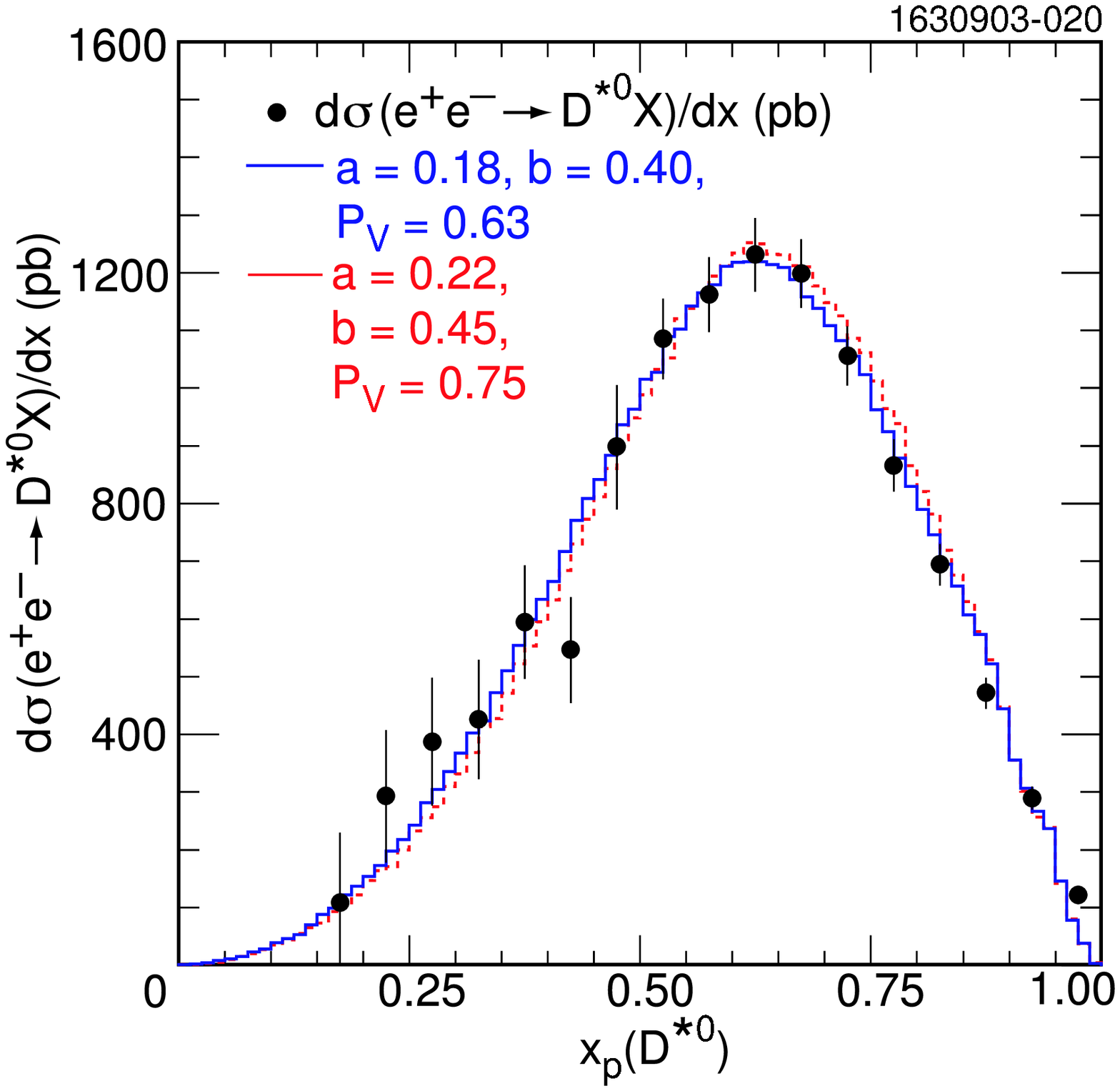}
\vspace{-0.2in}
\caption{\label{fig:DSZaver}$d\sigma(e^+e^-\to D^{*0}X)/dx_p$, weighted
average of the \DSZ\TO\DZ\PIZ\TO(\KM\PIP)\PIZ\ and
\DSZ\TO\DZ\PIZ\TO(\KM\PIP\PIP\PIM)\PIZ decay modes. Overlaid are the
JETSET spectra generated with two sets of parameters
(Sec.~\ref{sec:QQpar}).}\end{figure}

The weighted average of the spectra obtained
from the two decay modes analyzed is shown in Fig.~\ref{fig:DSZaver} and
listed in App.~\ref{sec:tables}, Table~\ref{tab:aver-sp}.
The two JETSET generated spectra are explained in Sec.~\ref{sec:QQpar}.

\section{Results for the Total Cross Sections and average $x_p$}
\label{sec:totcs}

The production cross section for each channel is shown in
Table~\ref{tab:totcs}.  It is calculated by summing each differential
cross section bin-by-bin. The first error in the table is the
statistical error, obtained by combining in quadrature the statistical
errors in each bin.  If the yield in the
lowest few bins cannot be reliably measured, the cross section
is corrected by extrapolating the spectrum to $x_p=0$ using the JETSET
distribution that fits the spectrum, discussed in Sec.~\ref{sec:QQpar}.
This correction is between 0.2\% and 6\%.

In Table~\ref{tab:syserr} we list, channel by channel, the components of
the systematic error on the production cross sections.  In the
first column we report the rms spread of the cross sections
obtained by the four or five smoothing sequences used for each channel. 
The discrepancy between the areas of the 
input and reconstructed spectra in the generic \MC\ check 
(Sec.~\ref{sec:checks}), is shown in the second column.  In the
third column we list the percent difference between the integral of the
spectra obtained using the double Gaussian and the one that uses the
TAGMC signal shape (Sec.~\ref{sec:proc}).  This error is not considered
for the channels where the \DZ\ decays to \KM\PIP\PIP\PIM, because of
the use of the COUNT procedure for those channels.  We assume a 10\%
error on the extrapolation and show it in column 4.  The remaining
systematic errors are estimated and discussed in a series of CLEO
internal notes and are used in all CLEO analyses where they are
relevant.  We estimate a 1\% per track uncertainty in the charged-track
detection efficiency and 0.8\% per track for particle identification
efficiency.  The choice of track quality and geometrical cuts result in
an error of 0.5\% also per track. The per track errors, being coherent,
are multiplied by the number of tracks in the decay, and are shown in
columns 5, 6, and 7.  The $\pi^0$ detection uncertainty is estimated to
be 3\% per $\pi^0$ (column 8).  
As discussed in 
Sec.~\ref{sec:errors}, we attribute a 0.5\% error due to possible
inaccuracies in the \MCS\ of the initial state radiation.  The error on
the integrated luminosity is estimated as 1.9\%.

\begin{table*}[htb]
\caption{\label{tab:syserr}Systematic errors described in the
text. Some are listed as percent of the cross section, other ones
directly in pb.  The momentum dependent systematic errors are listed
also in the tables in App.~\ref{sec:tables}.  The error due to the
uncertainty on the branching ratio is shown only in
Table~\ref{tab:totcs}.}
\begin{center}\begin{tabular}{|l|c|c|c|c|c|c|c|c|c|c|}\hline
                                             &    1     &      2      &  3   &   4   &    5   &    6    &   7   &      8    &   9    &  10   \\
                                             &procedures&     gMC     &signal&Extra- & track  &\ part.\ & other &\ $\pi^0$\ &\ ISR\  &       \\ 
Decay channel                                &   rms    &\ \ check\ \ &shape &polat. &det.eff.& ID      &  sel. &    det.   &\ sim.\ & Lum.\ \\ \hline 
$D^+\rightarrow K^-\pi^+\pi^+$               &\ 5pb     & 15pb        & 1.6\%& 0.5pb & 3\%    & 2.4\%   & 1.5\% &           & 0.5\%  & 1.9\%  \\
$D^0\rightarrow K^-\pi^+$         	     & 22pb 	&\ 8pb 	      & 1.6\%& 0.4pb & 2\%    & 1.6\%   & 1.0\% &           & 0.5\%  & 1.9\%  \\
$D^0\rightarrow K^-\pi^+\pi^+\pi^-$	     & 41pb 	& 29pb 	      &      & 3.2pb & 4\%    & 3.2\%   & 2.0\% &           & 0.5\%  & 1.9\%  \\
$D^{*+}\rightarrow (K^-\pi^+)\pi^+$          &\ 8pb 	& 15pb 	      & 1.6\%& 0.9pb & 3\%    & 2.4\%   & 1.5\% &           & 0.5\%  & 1.9\%  \\
$D^{*+}\rightarrow (K^-\pi^+\pi^+\pi^-)\pi^+$& 17pb 	&\ 7pb 	      &      & 3.3pb & 5\%    & 4.0\%   & 2.5\% &           & 0.5\%  & 1.9\%  \\
$D^{*0}\rightarrow (K^-\pi^+)\pi^0$          & 11pb 	& 10pb 	      & 1.6\%& 3.6pb & 2\%    & 1.6\%   & 1.0\% & 3\%       & 0.5\%  & 1.9\%  \\
$D^{*0}\rightarrow (K^-\pi^+\pi^+\pi-)\pi^0$ & 45pb 	& 12pb 	      &      & 1.1pb & 4\%    & 3.2\%   & 2.0\% & 3\%       & 0.5\%  & 1.9\%  \\
\hline\end{tabular}\end{center}\end{table*}

These systematic errors are combined in quadrature to give the
systematic error on the cross section, the second entry in
Table~\ref{tab:totcs}.

\begin{table*}[htb] 
\caption{\label{tab:totcs}Total production cross sections and average
\xp, as derived from each decay mode. The cross section errors are, in
this order, the statistical error, the systematic error and the error
due to the uncertainty on the branching ratio.}
\begin{center}\begin{tabular}{|l|c|}\hline
Decay channel & Total Cross Section (pb) at 10.5~GeV C.M.E.\\ \hline
$D^+\rightarrow K^-\pi^+\pi^+$&$\sigma(\ee\to D^+ X)=640\PM14\PM35\PM43$\\

$D^0\rightarrow K^-\pi^+$&$\sigma(\ee\to D^0 X)=1,521\PM16\PM62\PM36$\\ 

$D^0\rightarrow K^-\pi^+\pi^+\pi^-$&\ $\sigma(\ee\to D^0 X)=
1,579\PM55\PM102\PM63$\ \\

$D^{*+}\rightarrow D^0\pi^+\rightarrow (K^-\pi^+)\pi^+$&$\sigma(\ee\to
D^{*+} X)= 583\PM8\PM33\PM14$ \\

$D^{*+}\rightarrow D^0\pi^+\rightarrow (K^-\pi^+\pi^+\pi^-)\pi^+$\ &
$\sigma(\ee\to D^{*+} X)= 572\PM26\PM45\PM24$ \\

$D^{*0}\rightarrow D^0\pi^0\rightarrow (K^-\pi^+)\pi^0$&$\sigma(\ee\to
D^{*0} X)= 559\PM24\PM35\PM29$ \\

$D^{*0}\rightarrow D^0\pi^0\rightarrow (K^-\pi^+\pi^+\pi-)\pi^0$&
$\sigma(\ee\to D^{*0} X)= 616\PM32\PM62\PM39$  \\
\hline\end{tabular}\end{center}\end{table*}

We calculate $<x_p>$ for the $D^+$ spectrum and for the spectra of $D^0$, 
$D^{*+}$ and $D^{*0}$ averaged over the decay modes.  We
supplement the data spectrum in the lowest bins using the JETSET spectra
normalized to the spectra.  We take the errors on these
``borrowed'' cross sections to be roughly comparable to the data in
nearby bins.  The results are shown in Table~\ref{tab:avex}.

\vspace{-0.2in}\begin{table*}[hbt] 
\caption{\label{tab:avex} $<x_p>$ for the four charm mesons
  considered.  The first error is statistical, the second systematic.}
\begin{center}\begin{tabular}{|c|c||c|c|}\hline
Meson & $<x_p>$ &  Meson & $<x_p>$ \\ \hline
$D^+$& $0.582\pm0.008\pm0.004$ & $D^{*+}$& $0.611\pm0.007\pm0.004$\\
$D^0$& $0.570\pm0.005\pm0.004$ & $D^{*0}$& $0.596\pm0.009\pm0.004$\\ 
\hline\end{tabular}\end{center}\end{table*}

\section{Optimization of JETSET parameters}\label{sec:QQpar}

Largely for internal use of our collaboration, we perform a simple fit
of the \DZ\ spectrum (from the \DZ\TO\KM\PIP\ decay mode) varying the
three JETSET parameters that are most important for the shape of the
spectrum.  The first and second are the parameters $a$ and $b$ appearing
in the ``Lund Symmetric Fragmentation
Function''~\cite{Boand,LundSM}:
\begin{equation}\label{LSFF}
f(z)=N\frac{(1-z)^a}{z} \exp\left[\frac{-b\cdot
m^2_\bot}{z}\right]\end{equation} where $z$ is the reduced energy $x_E$,
or momentum $x_p$, of the hadron and $m^2_\bot=m^2+p^2_\bot$, with $m$
being the hadron mass and $p_\bot$ the component of the hadron momentum
perpendicular to the jet axis.

The third parameter is the probability $P_V$ that a meson of given
flavor be generated as a vector meson, rather than pseudoscalar or
tensor, $P_V\equiv V/(P+V+T)$.  The data indicate, as expected, that the
majority of \DZ 's are not produced directly in the fragmentation of the
charm quark, but from the decay of \DS 's.  In JETSET~\cite{jtst74}
these parameters are PARJ(41), PARJ(42) and PARJ(13).

The result of the fit of the \DZ\ spectrum (in the \KM\PIP\ decay mode)
is: \\
\[a= 0.178\pm0.007,\hspace{0.4in} b = 0.393\PM0.006,\hspace{0.4in}
P_V = 0.627\pm0.015.\] 
Keeping $P_V$ fixed at the naive value $P_V=0.75$, we obtain $a =
0.223 \PM 0.009$ and $b = 0.438 \PM 0.005$.  In both cases the quoted
errors are simple statistical errors.  Correlation between parameters
are not evaluated.  The spectra resulting from
these parameterizations are shown in Fig.~\ref{fig:DPspectrum},\
~\ref{fig:DZaver},~\ref{fig:DSPaver},~\ref{fig:DSZaver}. 

Notice that we do not consider our results of \DP, \DSP\ and \DSZ\
spectra in the optimization process.  However, \textit{a posteriori} we
see, visually from the figures, that the spectra generated with these
parameters seem to reproduce rather accurately also the \DP, \DSP\ and
\DSZ\ experimental distributions.  However, it is not obvious which one
of the two sets, the one with $P_V=0.672$ or the one with $P_V=0.75$,
should be preferred. Furthermore, these parameters, 
while useful for the \MC\ simulation of \D\ and \DS\ spectra at the
c.m. energy of our and similar experiments, should not be taken as having
general validity and theoretical significance.  In fact, the \Ds\
spectrum generated by JETSET with our fitted parameters disagrees
appreciably with the spectrum measured by the CLEO~\cite{EdJohn} and
BaBar~\cite{BabarDs} collaborations.  It should be noted that the
effect of these parameters may also be influenced by the value of other
JETSET parameters. 

\section{Conclusions}

We have measured the momentum distribution of \DZ, \DP, \DSP\ and \DSZ\
produced in non-resonant \ee\ annihilation at a CME of about 10.5~GeV.
These distributions can be used to guide and check QCD calculations of
fragmentation functions needed to predict heavy meson production in both
\ee\ annihilation and hadron collisions at very high energy.  The \DZ\
spectrum was used to determine the JETSET parameters that best reproduce
it, and we found that, with these parameters, the \DS, \DSP\ and
\DSZ\ spectra (but not the \Ds\ spectrum) are also well reproduced.

\begin{acknowledgments}
We gratefully acknowledge the effort of the CESR staff in providing us
with excellent luminosity and running conditions.  G. Moneti thanks
M. Cacciari and P. Nason for very useful discussions
on QCD calculations of heavy flavour fragmentation.  M. Selen thanks the
Research Corporation, and A.H. Mahmood thanks the Texas Advanced
Research Program.  This work was supported by the National Science
Foundation and the U.S. Department of Energy.
\end{acknowledgments}


\appendix
\section{Plots of detection efficiencies vs $x_p$}\label{sec:eff.plots}
In the following figures we show the detection efficiency dependence on
\xp\ for all the mesons and decay modes analyzed.  The detection
efficiencies obtained from the signal and generic MC\ simulations are
plotted, together with the curve resulting from the fit of their
weighted average to a polynomial.

\begin{figure*}[htb]\center
\includegraphics*[width=5.6in]{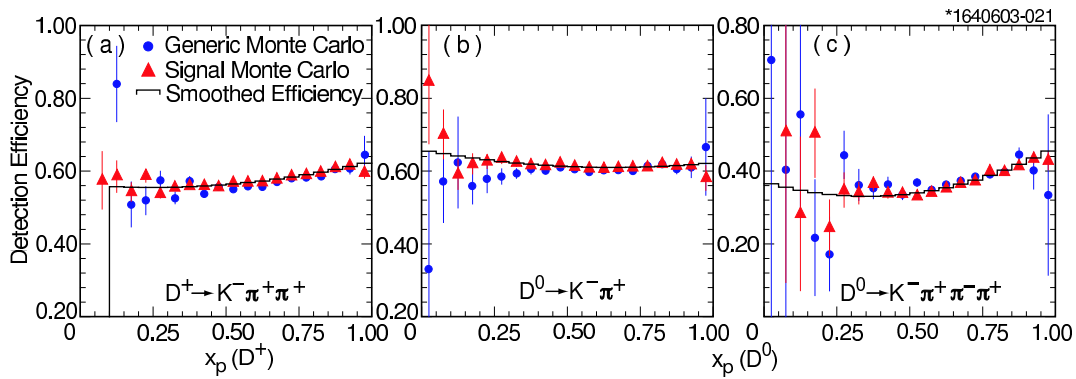}
\vspace{-0.2in}
\caption{\label{fig:dp6-gmct}Direct comparison of the detection
efficiencies from signal and generic \MC\ and the result of smoothing
their average: (a) for the \DP\TO\KM\PIP\PIP\ channel, (b) for the
\DZ\TO\KM\PIP\ channel, and (c) for the \DZ\TO\KM\PIP\PIP\PIM\
channel.}\end{figure*}

\begin{figure*}[h!tb]\center
\includegraphics*[width=3.8in]{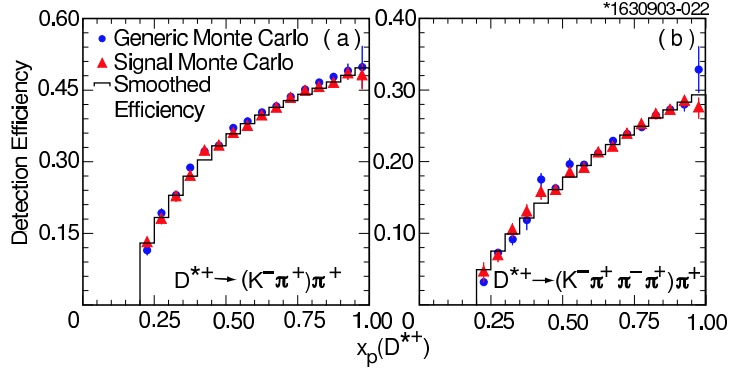}
\vspace{-0.25in}
\caption{\label{fig:dspz13-gmct}Comparison of the detection efficiencies
obtained from the signal and generic \MC\ and their smoothed average:
(a) for the $D^{*+}\rightarrow D^0\pi^+\rightarrow (K^-\pi^+)\pi^+$
channel, (b) for the $D^{*+}\rightarrow D^0\pi^+\rightarrow
(K^-\pi^+\pi^-\pi^+)\pi^+$ channel.}\end{figure*}

\begin{figure*}[h!tb]
\includegraphics*[width=3.8in]{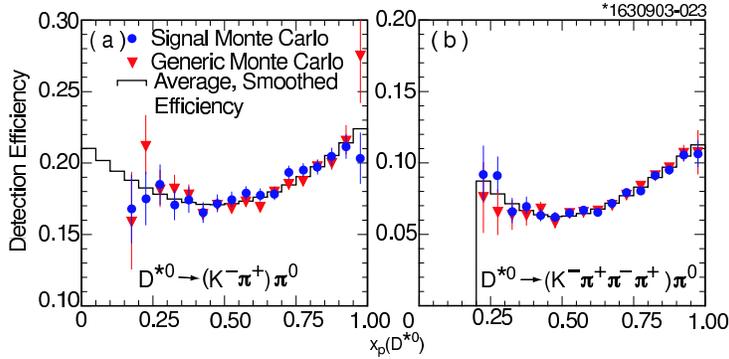}
\vspace{-0.2in}
\caption{\label{fig:dszz13-gmct}Comparison of the
unsmoothed detection efficiencies obtained from the signal and generic
\MC: (a) for the $D^{*0}\rightarrow D^0\pi^0\rightarrow
(K^-\pi^+)\pi^0$, (b) for the $D^{*0}\rightarrow
D^0\pi^0\rightarrow(K^-\pi^+\pi^-\pi^+)\pi^0$ channel.}\end{figure*}

\section{Tables of differential cross sections} \label{sec:tables}

In the following tables, we report the quantity $d\sigma/dx_p$ in pb.
Notice that the systematic and total errors are errors on the bin
content (\ie, the first column).  The first column of systematic errors
is obtained from the rms spread of yields for the different procedures
used to calculate the spectrum.  The second column of systematic errors
is derived from the ``generic MC check'' described in
Sec.~\ref{sec:checks}. 
These are the errors relevant to the
shape of the spectra, \ie, they do not include the systematic errors
that are common to the whole momentum range and that contribute to the
error on the cross section (Sec.~\ref{sec:totcs}).  

\begin{table}[thb] 
\caption{\label{tab:DPspectrum}$d\sigma(e^+e^-\to D^+ X)/dx_p$ in pb;
(\DP\TO\KM\PIP\PIP)}
\begin{center}\begin{tabular}{||c||c|c|c|c|c||}\hline\hline
   &\ \  \ $d\sigma/dx_p$\ \ \ & \multicolumn{4}{c||}{Errors (pb)}\\ 
\ \ \ \ \ \ $x_p$\ \ \ \ \ \ & (pb) & Statistical &
\multicolumn{2}{c|}{\ \ \ Systematic\ \ \ } &\ \ \ Total\ \ \ \  \\ \hline  
 0.15-0.20 &  161  &  78  &\ \ \ \ 27\ \ \ \ &\ \ \ \ 3\ \ \ \ \  &  83 \\
 0.20-0.25 &  320  &  76  &  53   &   5   & 92  \\
 0.25-0.30 &  356  &  70  &  59   &   6   & 92  \\
 0.30-0.35 &  413  &  64  &  68   &   7   & 94  \\
 0.35-0.40 &  693  &  58  &  11   &  11   & 60  \\
 0.40-0.45 &  909  &  52  &  14   &  15   & 56  \\
 0.45-0.50 & 1042  &  47  &  16   &  17   & 53  \\
 0.50-0.55 & 1271  &  25  &  20   &  21   & 38  \\
 0.55-0.60 & 1357  &  22  &  21   &  22   & 38  \\
 0.60-0.65 & 1370  &  19  &  21   &  22   & 36  \\
 0.65-0.70 & 1291  &  17  &  20   &  21   & 34  \\
 0.70-0.75 & 1129  &  15  &  17   &  18   & 29  \\
 0.75-0.80 &  952  &  13  &  15   &  16   & 25  \\
 0.80-0.85 &  694  &  10  &  11   &  11   & 19  \\
 0.85-0.90 &  449  &   8  &   7   &   7   & 13  \\
 0.90-0.95 &  223  &   5  &   3   &   4   &  7  \\
 0.95-1.00 &   74  &   3  &   1   &   1   &  4  \\
\hline\hline	      
\end{tabular}\end{center}\end{table} 

\begin{table}[thb]			
\caption{\label{tab:DZ1spectrum}$d\sigma(e^+e^-\to D^0 X)/dx_p$ in pb;
(\DZ\TO\KM\PIP).}
\begin{center}\begin{tabular}{||c||c|c|c|c|c||}\hline\hline
   &\ \  \ $d\sigma/dx_p$\ \ \ & \multicolumn{4}{c||}{Errors (pb)}\\ 
\ \ \ \ \ \ $x_p$\ \ \ \ \ \ & (pb) & Statistical &
\multicolumn{2}{c|}{\ \ \ Systematic\ \ \ } &\ \ \ Total\ \ \ \  \\ \hline  
 0.10-0.15  &  196  &  86  &\ \ \ \ 73\ \ \ \ &\ \ \ \ 1\ \ \ \ \ & 113 \\
 0.15-0.20  &  507  &  92  & 188  &   3  & 209  \\
 0.20-0.25  &  597  &  85  & 221  &   3  & 237  \\
 0.25-0.30  &  891  &  76  &  37  &   5  &  85  \\
 0.30-0.35  & 1154  &  68  &  48  &   7  &  84  \\
 0.35-0.40  & 1665  &  63  &  70  &  10  &  95  \\
 0.40-0.45  & 2341  &  61  &  98  &  13  & 116  \\
 0.45-0.50  & 2889  &  59  & 121  &  17  & 136  \\
 0.50-0.55  & 3178  &  35  &  42  &  18  &  57  \\
 0.55-0.60  & 3444  &  34  &  45  &  20  &  60  \\
 0.60-0.65  & 3345  &  34  &  44  &  19  &  58  \\
 0.65-0.70  & 2984  &  33  &  39  &  17  &  54  \\
 0.70-0.75  & 2542  &  31  &  33  &  15  &  48  \\
 0.75-0.80  & 1997  &  29  &  26  &  11  &  41  \\
 0.80-0.85  & 1380  &  25  &  18  &   8  &  32  \\
 0.85-0.90  &  831  &  19  &  11  &   5  &  23  \\
 0.90-0.95  &  337  &  11  &   4  &   2  &  12  \\
 0.95-1.00  &   78  &   5  &   1  &  0.4 &   6  \\
\hline\hline\end{tabular}\end{center}\end{table} 
					        
\begin{table}[thp]			        
\caption{\label{tab:DZ3spectrum}$d\sigma(e^+e^- \to D^0 X)/dx_p$ in pb;
  (\DZ\TO\KM\PIP\PIP\PIM).}		        
\begin{center}\begin{tabular}{||c||c|c|c|c|c||} \hline\hline
   &\ \  \ $d\sigma/dx_p$\ \ \ & \multicolumn{4 }{c||}{Errors (pb)}\\ 
\ \ \ \ \ \ $x_p$\ \ \ \ \ \ & (pb) & Statistic al &
\multicolumn{2}{c|}{\ \ \ Systematic\ \ \ } &\  \ \ Total\ \ \ \  \\ \hline  
 0.15-0.20  &  146  &  283  &\ \ \ 291\ \ \ \ & \ \ \ \ 4\ \ \ \ \ & 406 \\
 0.20-0.25  &  292  &  430  &  101  &    9  &  441  \\
 0.25-0.30  &  551  &  481  &  190  &   16  &  518  \\
 0.30-0.35  & 1343  &  525  &  464  &   40  &  702  \\
 0.35-0.40  & 2068  &  479  &  715  &   61  &  862  \\
 0.40-0.45  & 2420  &  323  &   60  &   72  &  337  \\
 0.45-0.50  & 2552  &  254  &   63  &   76  &  272  \\
 0.50-0.55  & 3500  &  211  &   86  &  104  &  250  \\
 0.55-0.60  & 3868  &  151  &   95  &  115  &  212  \\
 0.60-0.65  & 3651  &  127  &   90  &  108  &  190  \\
 0.65-0.70  & 3274  &  134  &   81  &   97  &  184  \\
 0.70-0.75  & 2635  &  143  &   65  &   78  &  175  \\
 0.75-0.80  & 2108  &   93  &   52  &   63  &  123  \\
 0.80-0.85  & 1403  &   71  &   35  &   42  &   89  \\
 0.85-0.90  &  815  &   49  &   20  &   24  &   59  \\
 0.90-0.95  &  355  &   27  &    9  &   11  &   31  \\
 0.95-1.00  &   87  &   12  &    2  &    3  &   13  \\
\hline\hline\end{tabular}\end{center}\end{table} 

\begin{table}[hbt]
\caption{\label{tab:DSPZ1spectrum}$d\sigma(e^+e^-\to D^{*+} X)/dx_p$ in
  pb; (\DSP\TO(\KM\PIP)\PIP).}
\begin{center}\begin{tabular}{||c||c|c|c|c|c||}\hline\hline
   &\ \  \ $d\sigma/dx_p$\ \ \ & \multicolumn{4}{c||}{Errors (pb)}\\ 
\ \ \ \ \ \ $x_p$\ \ \ \ \ \ & (pb) & Statistical &
\multicolumn{2}{c|}{\ \ \ Systematic\ \ \ } &\ \ \ Total\ \ \ \  \\ \hline  
 0.20-0.25  &  169   &  66  &\ \ \ 65\ \ \ & \ \ \ \ 1\ \ \ \ \ &  93 \\
 0.25-0.30  &  258   &  56  &   27  &   2  &  63 \\
 0.30-0.35  &  355   &  50  &   38  &   3  &  63 \\
 0.35-0.40  &  501   &  48  &   53  &   4  &  72 \\
 0.40-0.45  &  617   &  49  &   12  &   5  &  50 \\
 0.45-0.50  &  915   &  52  &   18  &   7  &  55 \\
 0.50-0.55  & 1103   &  30  &   22  &   9  &  38 \\
 0.55-0.60  & 1256   &  31  &   25  &  10  &  41 \\
 0.60-0.65  & 1293   &  31  &   25  &  10  &  41 \\
 0.65-0.70  & 1267   &  31  &   25  &  10  &  41 \\
 0.70-0.75  & 1125   &  30  &   22  &   9  &  38 \\
 0.75-0.80  &  947   &  29  &   19  &   7  &  35 \\
 0.80-0.85  &  731   &  26  &   14  &   6  &  30 \\
 0.85-0.90  &  529   &  22  &   10  &   4  &  25 \\
 0.90-0.95  &  303   &  16  &    6  &   2  &  17 \\
 0.95-1.00  &  116   &   9  &    2  &   1  &   9 \\
\hline\hline\end{tabular}\end{center}\end{table}

\begin{table}[thb]		
\caption{\label{tab:DSPZ3spectrum}$d\sigma(e^+e^-\to D^{*+} X)/dx_p$ in
pb; (\DSP\TO(\KM\PIP\PIP\PIM)\PIP).}
\begin{center}\begin{tabular}{||c||c|c|c|c|c||}\hline\hline
   &\ \  \ $d\sigma/dx_p$\ \ \ & \multicolumn{4}{c||}{Errors (pb)}\\ 
\ \ \ \ \ \ $x_p$\ \ \ \ \ \ & (pb) & Statistical &
\multicolumn{2}{c|}{\ \ \ Systematic\ \ \ } &\ \ \ Total\ \ \ \  \\ \hline  
 0.25-0.30  &  201   &  147  &\ \ \ 136\ \ \ &\ \ \ \ 4\ \ \ \ \ & 200 \\
 0.30-0.35  &  265   &  120  & 179 &   5  &  216 \\
 0.35-0.40  &  478   &  102  &  45 &   9  &  112 \\
 0.40-0.45  &  657   &   88  &  61 &  12  &  108 \\
 0.45-0.50  &  943   &   80  &  88 &  17  &  120 \\
 0.50-0.55  & 1121   &   45  &  27 &  20  &   57 \\
 0.55-0.60  & 1221   &   41  &  29 &  22  &   55 \\
 0.60-0.65  & 1276   &   36  &  30 &  23  &   52 \\
 0.65-0.70  & 1221   &   32  &  29 &  22  &   49 \\
 0.70-0.75  & 1096   &   29  &  26 &  20  &   44 \\
 0.75-0.80  &  915   &   25  &  22 &  17  &   37 \\
 0.80-0.85  &  715   &   21  &  17 &  13  &   30 \\
 0.85-0.90  &  533   &   18  &  13 &  10  &   24 \\
 0.90-0.95  &  317   &   14  &   8 &   6  &   17 \\
 0.95-1.00  &  122   &   10  &   3 &   2  &   11 \\
\hline\hline\end{tabular}\end{center}\end{table} 

\begin{table}[hbt]		
\caption{\label{tab:DSZZ1spectrum}$d\sigma(e^+e^-\to D^{*0} X)/dx_p$ in pb;
(\DSZ\TO(\KM\PIP)\PIZ).} 
\begin{center}\begin{tabular}{||c||c|c|c|c|c||}\hline\hline
   &\ \  \ $d\sigma/dx_p$\ \ \ & \multicolumn{4}{c||}{Errors (pb)}\\ 
\ \ \ \ \ \ $x_p$\ \ \ \ \ \ & (pb) & Statistical &
\multicolumn{2}{c|}{\ \ \ Systematic\ \ \ } &\ \ \ Total\ \ \ \  \\ \hline  
 0.15-0.20  &  108 &  121  &\ \ \ \ \ 6\ \ \ \ \ &\ \ \ \ 2\ \ \ \ \ &  121 \\
 0.20-0.25  &  290 &  121  &   17  &   7  &   123 \\
 0.25-0.30  &  376 &  112  &   23  &   9  &   114 \\
 0.30-0.35  &  425 &  104  &   26  &  10  &   107 \\
 0.35-0.40  &  580 &   95  &   35  &  13  &   102 \\
 0.40-0.45  &  601 &   92  &   36  &  14  &   100 \\
 0.45-0.50  &  946 &   99  &   57  &  22  &   116 \\
 0.50-0.55  & 1061 &   69  &   30  &  24  &    79 \\
 0.55-0.60  & 1124 &   61  &   31  &  26  &    73 \\
 0.60-0.65  & 1186 &   60  &   33  &  27  &    73 \\
 0.65-0.70  & 1125 &   56  &   31  &  26  &    69 \\
 0.70-0.75  &  992 &   48  &   28  &  23  &    60 \\
 0.75-0.80  &  822 &   47  &   23  &  19  &    55 \\
 0.80-0.85  &  662 &   36  &   18  &  15  &    43 \\
 0.85-0.90  &  425 &   28  &   12  &  10  &    32 \\
 0.90-0.95  &  271 &   24  &    8  &   6  &    26 \\
 0.95-1.00  &  107 &   22  &    3  &   2  &    22 \\
\hline\hline\end{tabular}\end{center}\end{table} 

\begin{table}[thb]		
\caption{\label{tab:DSZZ3spectrum}$d\sigma(e^+e^-\to D^{*0} X)/dx_p$ in pb;
(\DSZ\TO(\KM\PIP\PIP\PIM)\PIZ).} 
\begin{center}\begin{tabular}{||c||c|c|c|c|c||}\hline\hline
   &\ \  \ $d\sigma/dx_p$\ \ \ & \multicolumn{4}{c||}{Errors (pb)}\\ 
\ \ \ \ \ $x_p$\ \ \ \ \ & (pb) & Statistical &
\multicolumn{2}{c|}{\ \ \ Systematic\ \ \ } &\ \ \ Total\ \ \ \  \\ \hline  
\ \ \ 0.20-0.25\ \ \ & 308   &   251  &\ \ \ 206\ \ \ &
\ \ \ 7\ \ \ &  325 \\
 0.25-0.30  &  559   &   262  &   374  &   12  &  457 \\
 0.30-0.35  &  428   &   259  &   286  &    9  &  386 \\
 0.35-0.40  &  755   &   247  &   250  &   16  &  352 \\
 0.40-0.45  &  236   &   223  &    78  &    5  &  236 \\
 0.45-0.50  &  601   &   205  &   199  &   13  &  286 \\
 0.50-0.55  & 1173   &   135  &    64  &   25  &  152 \\
 0.55-0.60  & 1300   &   118  &    71  &   28  &  141 \\
 0.60-0.65  & 1367   &   100  &    75  &   29  &  128 \\
 0.65-0.70  & 1418   &    85  &    78  &   30  &  119 \\
 0.70-0.75  & 1235   &    70  &    68  &   27  &  101 \\
 0.75-0.80  &  954   &    56  &    52  &   21  &   80 \\
 0.80-0.85  &  764   &    46  &    42  &   16  &   64 \\
 0.85-0.90  &  581   &    36  &    32  &   12  &   50 \\
 0.90-0.95  &  317   &    26  &    17  &    7  &   32 \\
 0.95-1.00  &  131   &    18  &     7  &    3  &   20 \\
\hline\hline\end{tabular}\end{center}\end{table} 

\begin{table}[thb]
\caption{\label{tab:aver-sp}Differential cross sections $d\sigma/dx_p$ in
pb for $D^+$, $D^0$, $D^{*+}$ and $D^{*0}$. The last three columns are
weighted averaged over the two decay modes. The errors are the quadratic
combination of the statistical and systematic errors, excluding the
errors, discussed in Sec.~\ref{sec:totcs}, that affect the total cross
section but not the shape of the spectrum.}
 
\begin{center}\begin{tabular}
{|| c || c || c || c || c ||}\hline\hline
\ \ \ \ \ \ \ $x_p$\ \ \ \ \ \ &\ \ \ \ \ \ \ $D^+$\ \ \ \ \ \ \ &\ \ \
\ \ \ \ $D^0$\ \ \ \ \ \ \ &\ \ \ \ \ \ $ D^{*+}$\ \ \ \ \ \ &\ \ \ \ \
\ $D^{*0}$\ \ \ \ \ \ \\ \hline   
 0.10-0.15 &       -      &   173  \PM  109 &       -     &        -      \\
 0.15-0.20 &  161 \PM  83 &   431  \PM  186 &       -     &  108  \PM 121 \\
 0.20-0.25 &  320 \PM  92 &   529  \PM  209 &  146 \PM 86 &  292  \PM 115 \\
 0.25-0.30 &  356 \PM  92 &   882  \PM   84 &  253 \PM 60 &  387  \PM 111 \\
 0.30-0.35 &  413 \PM  94 &  1156  \PM   83 &  348 \PM 60 &  425  \PM 103 \\
 0.35-0.40 &  693 \PM  60 &  1670  \PM   94 &  494 \PM 60 &  594  \PM  98 \\
 0.40-0.45 &  909 \PM  56 &  2349  \PM  110 &  624 \PM 46 &  546  \PM  92 \\
 0.45-0.50 & 1042 \PM  53 &  2822  \PM  122 &  920 \PM 50 &  897  \PM 108 \\
 0.50-0.55 & 1271 \PM  38 &  3194  \PM   56 & 1108 \PM 32 & 1085  \PM  70 \\
 0.55-0.60 & 1357 \PM  38 &  3475  \PM   58 & 1244 \PM 33 & 1162  \PM  65 \\
 0.60-0.65 & 1370 \PM  36 &  3371  \PM   56 & 1286 \PM 32 & 1230  \PM  64 \\
 0.65-0.70 & 1291 \PM  34 &  3007  \PM   51 & 1248 \PM 31 & 1198  \PM  60 \\
 0.70-0.75 & 1129 \PM  29 &  2549  \PM   46 & 1113 \PM 29 & 1055  \PM  52 \\
 0.75-0.80 &  952 \PM  25 &  2008  \PM   39 &  932 \PM 25 &  865  \PM  45 \\
 0.80-0.85 &  694 \PM  19 &  1383  \PM   30 &  723 \PM 21 &  694  \PM  36 \\
 0.85-0.90 &  449 \PM  13 &   829  \PM   21 &  531 \PM 17 &  471  \PM  27 \\
 0.90-0.95 &  223 \PM   7 &   339  \PM   11 &  310 \PM 12 &  289  \PM  20 \\
 0.95-1.00 &   74 \PM   4 &    90  \PM    5 &  119 \PM  7 &  121  \PM  15 \\
\hline\hline\end{tabular}\end{center}\end{table} 


\begin{thebibliography}{99}

\bibitem{Alexander:1995aj}
G.~Alexander {\it et al.}  [OPAL Collaboration], Phys.\ Lett.\ {\bf
  B364}, 93 (1995).

\bibitem{Abe:1999ki} K.~Abe {\it et al.}  [SLD Collaboration], Phys.\
Rev.\ Lett.\ {\bf 84}, 4300 (2000) [arXiv:hep-ex/9912058].

\bibitem{Heister:2001jg} A.~Heister {\it et al.}  [ALEPH Collaboration],
Phys.\ Lett.\ {\bf B512}, 30 (2001) [arXiv:hep-ex/0106051].

\bibitem{Adeva:1991iw} B.~Adeva {\it et al.}  [L3 Collaboration], Phys.\
Lett.\ {\bf B261}, 177 (1991). 

\bibitem{Argus91} H. Albrecht \etal\ [ARGUS Collaboration],
  Z. Phys. {\bf C52}, 353 (1991). 

\bibitem{CLEO88} D. Bortoletto \etal\ [CLEO Collaboration],
  Phys. Rev. {\bf D37}, 1719 (1988).

\bibitem{biebel} O. Biebel, P. Nason, B.R. Webber [arXiv:hep-ph/0109282]
v2, an abbreviated version is in~\cite{PDG}.

\bibitem{Nason:1993xx} P.~Nason and B.~R.~Webber, Nucl. Phys. {\bf
  B421}, 473 (1994) [Erratum-ibid.\ {\bf B480}, 755 (1996)].

\bibitem{Ben-Haim} E. Ben-Haim \etal, [arXiv:hep-ph/0302157].

\bibitem{Cacciari1} M. Cacciari and S. Catani,  Nucl. Phys. {\bf B617},
  253 (2001) [arXiv:hep-ph/0107138].

\bibitem{Abbott} B. Abbott \etal\ [D0 Collaboration], Phys. Lett. {\bf
B487}, 264 (2000). 

\bibitem{Acosta-b}  D. Acosta \etal\ [CDF Collaboration],
Phys. Rev. {\bf D65}, 052005 (2002).

\bibitem{Acosta-c} D. Acosta \etal\ [CDF Collaboration],
Phys. Rev. Lett. {\bf 91}, 241804 (2003) [arXiv:hep-ex/0307080].

\bibitem{H1-c} S. Aid \etal\ [H1 Collaboration], Z. Phys. {\bf C72}, 593
  (1996). 

\bibitem{ZEUS-c} S. Chekanov \etal\ [ZEUS Collaboration],
  [arXiv:hep-ex/0308068]. 

\bibitem{Cacciari2} M. Cacciari and E. Gardi, Nucl.\ Phys.\ {\bf B664},
  299 (2003) [arXiv:hep-ph/0301047].

\bibitem{Altarelli:1977zs} G.~Altarelli and G.~Parisi, Nucl.\ Phys.\ 
{\bf B126}, 298 (1977). 

\bibitem{Furmanski:1980cm} W.~Furmanski and R.~Petronzio, Phys.\ Lett.\
{\bf B97}, 437 (1980). 

\bibitem{ALEPH99} R. Barate \etal\ [ALEPH Collaboration],  
Eur. Phys. J. {\bf C16}, 597-611 (2000) [arXiv:hep-ex/9909032]. 

\bibitem{Artru} X. Artru, G. Menessier, Nucl. Phys. {\bf 70}, 93 (1974).

\bibitem{Boand} Bo Andersson \etal, Phys. Reports {\bf 97}, 31 (1983).

\bibitem{LundSM} Bo Andersson, ``The Lund Model'', Cambridge U. Press (1998).

\bibitem{Marchesini} G. Marchesini \etal, Comp. Phys. Comm. {\bf 67},
  465 (1992); G. Corcella \etal, JHEP {\bf 0101}, 010 (2001).

\bibitem{jtst74} T. Sjostrand, Comp. Phys. Comm. {\bfseries 82}, 74-89,
(1994), T. Sjostrand, ``PYTHIA 5.7 and JETSET 7.4 Physics and Manual''
[hep-ph/9508391].

\bibitem{Chun} S. Chun and C. Buchanan, Phys. Reports {\bf 292}, 239 (1998).

\bibitem{cleo1997} L. Gibbons \etal\ [CLEO Collaboration], Phys. Rev. {\bf
D56}, 3783 (1997).

\bibitem{cleo2} Y. Kubota \etal\ [CLEO Collaboration],
Nucl. Instrum. Methods Phys. Res., Sec. A {\bf 320}, 66(1992).

\bibitem{silicon} T. Hill \etal\ [CLEO Collaboration],
Nucl.  Instrum. Methods Phys. Res., Sec. A {\bf 418}, 32(1998).

\bibitem{Petra}HRS Collaboration, M.Derrick \etal\ Phys. Lett. {\bf
  246B}, 261, (1984); TPC/Two-Gamma Collaboration, H. Aihara \etal\
  Phys. ReV. {\bf D34}, 1945 (1986); TASSO Collaboration, M. Althoff
  \etal\ Phys. Lett. {\bf 126B}, 493 (1983); JADE Collaboration,
  W. Bartel \etal\ Phys. Lett. {\bf 161B}, 197 (1985).

\bibitem{SJO}
T. Sj\"ostrand, Comp. Phys. Comm. {\bf 39}, 347 (1986), T. 
Sj\"ostrand, M.~Bengston, Comp. Phys. Comm. {\bf 43}, 367 (1987).

\bibitem{GEANT} R. Brun \etal, ``GEANT, Detector Description and
  Simulation Tool'', CERN Program Library Long Writeup W5013, 1993.

\bibitem{PDG}
 D. E. Groom \etal\ [PDG], Eur. Phys. Jour. {\bf 15}, 1, (2000), 
 K. Hagiwara \etal\ [PDG],\Journal{\PRD},{66},{010001},{2002}, and 
 L. Alvarez-Gaume' \etal\ [PDG] Phys. Lett. {\bf B592}, 1, (2004). 

\bibitem{D*align} G. Branderburg \etal\ [CLEO Collaboration]
  Phys. Rev. {\bf D58}, 052003 (1998).

\bibitem{EdJohn} R. A. Briere \etal\ [CLEO Collaboration],
  Phys. Rev. {\bf D62}, 072003 (2000).

\bibitem{BabarDs} B. Aubert \etal\ [BABAR Collaboration],
  Phys. Rev. {\bf D65} (2002) 091104 [hep-ex/0201041].

\end{thebibliography}
\end{document}